\documentclass[12pt]{article}
\pdfoutput=1
\usepackage{amssymb}
\usepackage{amsmath}
\usepackage{latexsym}
\usepackage[usenames]{color}
\usepackage{cite}
\usepackage{setspace}
\usepackage{mathtools}
\usepackage{simplewick}
\numberwithin{equation}{section}
\definecolor{darkblue}{cmyk}{0.9,0.9,0,0}
\definecolor{darkred}{rgb}{0.6,0,0.3}
\usepackage{graphicx} 
\usepackage{epsfig}
\usepackage[setpagesize=false,pagebackref=false, linktocpage, bookmarksopen=true, colorlinks=true, linkcolor=darkblue,citecolor=darkblue,urlcolor=darkblue]{hyperref}
\usepackage{hyperref}
\newcommand{\arXiv}[2]{\href{http://arxiv.org/abs/#1}{{\tt #2}}}
\newcommand{\hep}[2]{\href{http://arxiv.org/abs/#1}{{\tt #2}}}

\newcommand{\tr}{{\rm tr}}

\def\del{\partial}

\def\fn#1{\footnote{#1}}
\def\nn{\nonumber}
\def\eqref#1{(\ref{#1})}
\def\comma{\,,}
\def\period{\,.}
\def\dmatrix#1#2{\left| 
\begin{array}{#1}
#2\end{array} 
\right|}
\def\NO#1{:\!#1\!:\,}
\newcommand{\beq}{\begin{equation}}
\newcommand{\eeq}{\end{equation}}

\begin{document}
\thispagestyle{empty}

\renewcommand{\thefootnote}{\fnsymbol{footnote}}
\setcounter{page}{1}
\setcounter{footnote}{0}
\setcounter{figure}{0}
\begin{flushright}
PUTP-2549 \\
\end{flushright}
\begin{center}
$$$$
{\large\textbf{\mathversion{bold}
{Exact Correlators on the Wilson Loop in ${\cal N}=4$ SYM:\\
Localization, Defect CFT, and Integrability}}\par}

\vspace{1.6cm}

\textrm{Simone Giombi\fn{\tt sgiombi AT princeton.edu}, Shota Komatsu\fn{\tt skomatsu AT ias.edu}}
\\ \vspace{2cm}
\footnotesize{\textit{
$^{\ast}$ Department of Physics, Princeton University, Princeton, NJ 08544, USA
\vspace{1mm} \\
$^{\dagger}$ School of Natural Sciences, Institute for
Advanced Study, Princeton, NJ 08540, USA
}  
\vspace{4mm}
}

\par\vspace{2.5cm}

\textbf{Abstract}\vspace{2mm}
\end{center}
We compute a set of correlation functions of operator insertions on the $1/8$ BPS Wilson loop 
in $\mathcal{N}=4$ SYM by employing supersymmetric localization, OPE and the Gram-Schmidt orthogonalization. 
These correlators exhibit a simple determinant structure, are position-independent and form a topological subsector, 
but depend nontrivially on the 't~Hooft coupling and the rank of the gauge group. 
When applied to the $1/2$ BPS circular (or straight) Wilson loop, our results provide an infinite family of exact defect CFT data, 
including the structure constants of protected defect primaries of arbitrary length inserted on the loop. At strong coupling, we show 
precise agreement with a direct calculation using perturbation theory around the AdS$_2$ string worldsheet. 
We also explain the connection of our results to the ``generalized Bremsstrahlung functions" previously computed from 
integrability techniques, reproducing the known results in the planar limit as well as obtaining some of their finite $N$ generalizations. 
Furthermore, we show that the correlators at large $N$ can be recast as simple integrals of products of polynomials (known as $Q$-functions)
that appear in the Quantum Spectral Curve approach. This suggests an interesting interplay between localization, defect CFT and integrability.
\noindent

\setcounter{page}{1}
\renewcommand{\thefootnote}{\arabic{footnote}}
\setcounter{footnote}{0}
\setcounter{tocdepth}{2}
\newpage
\tableofcontents

\parskip 5pt plus 1pt   \jot = 1.5ex

\newpage
\section{Introduction\label{sec:intro}}
The exact solution to an interacting quantum field theory in four dimensions would mark a breakthrough in theoretical physics, although it still seems out of reach at present time. In supersymmetric theories, one can make some progress since there are observables that preserve a fraction of the supersymmetries and are therefore often amenable to exact analytic methods, most notably supersymmetric localization \cite{Pestun1}. 

Another powerful method, which is currently the subject of active exploration, is the conformal bootstrap, see e.g. \cite{Simmons-Duffin:2016gjk} for a recent review. This approach uses conformal symmetry instead of supersymmetry, and has been remarkably successful in deriving bounds on physical quantities in non-trivial CFTs (most notably the 3d Ising model) and in charting a landscape of theories from a minimal set of assumptions \cite{RRTV,EPPRS}.

The third way towards this goal is integrability \cite{review}. Although the applicability of integrability is much smaller than the other two since it applies only to specific theories, the advantage is that it works not only for supersymmetric observables but for non-supersymmetric ones as well. It also allows one to compute them exactly as a function of coupling constants, rather than giving general bounds.

In this paper, we consider quantities which may stand at the ``crossroads'' of all these three methods. More specifically, we study the correlation functions of local operator insertions on the $1/8$-BPS Wilson loop in $\mathcal{N}=4$ supersymmetric Yang-Mills theory (SYM). 
The supersymmetric Wilson loop in $\mathcal{N}=4$ SYM has been an active subject of study since the early days of AdS/CFT correspondence \cite{Maldacena, ReyYee}. The $1/2$-BPS circular Wilson loop, which preserves a maximal amount of supersymmetry, was computed first by summing up a class of Feynman diagrams \cite{ESZ,DG}, and the exact result for its expectation value was later derived rigorously from supersymmetric localization \cite{Pestun1}, which reduces the problem to a simple Gaussian matrix model. The result is a nontrivial function of the coupling constant, which nevertheless matches beautifully with the regularized area of the string in AdS at strong coupling, providing key evidence for the holographic duality.

The computation was subsequently generalized to less supersymmetric Wilson loops, such as the $1/4$ BPS circular loop \cite{DrukkerQuarter}, and a more general infinite family of $1/8$-BPS Wilson loops defined on curves of arbitrary shape on a two-sphere \cite{DGRT1,DGRT2}. For such loops, an exact localization to 2d Yang-Mills theory was conjectured in \cite{DGRT1,DGRT2}, and later supported by a localization calculation in \cite{Pestun2}. Because of the invariance under area-preserving diffeomorphisms of 2d YM theory, one finds that the result for the expectation value of the $1/8$ BPS Wilson loop depends only on the area of the region surrounded by the loop. The localization relation to the 2d theory was checked in a number of non-trivial calculations, see e.g. \cite{Giombi:2009ms, Bassetto:2009rt, Giombi:2009ek, Bassetto:2009ms, GP1, GP2, Bonini:2014vta, BGPS}. It was also used in \cite{CHMS} to compute various important quantities defined on the Wilson loop, such as the two-point function of the displacement operator and the related ``Bremsstrahlung function''. It was based on the observation that one can insert a displacement operator by differentiating the expectation value of the Wilson loop with respect to its area $A$; $D\sim\del_{A} \langle\mathcal{W}\rangle$.

The purpose of this paper is to show that there are infinitely many other observables that can be computed using the results from localization. They are the correlation functions of special scalar insertions $\tilde{\Phi}^{L}$ inside the Wilson loop trace, where the scalar $\tilde{\Phi}$ is chosen so that the correlators are independent of the positions of the insertions.\footnote{The correlators of Wilson loops and local operators of similar kind, 
but inserted away from the Wilson loop, was studied in earlier literature, e.g. \cite{GP1, Bonini:2014vta, GP2}.} Similarly to the displacement operator, one can relate the insertion of $\tilde{\Phi}$'s to the area-derivative of the Wilson loop, essentially because $\tilde{\Phi}$ turns out to correspond via localization to insertions of the Hodge dual of the 2d YM field strength. However, one key difference from the analysis in \cite{CHMS} is that after taking the multiple area derivatives, to define the properly normal-ordered operators one has to perform the so-called Gram-Schmidt orthogonalization to make $\tilde{\Phi}^{k}$'s for different $k$ orthogonal to each other (and, in particular, also orthogonal to the identity, i.e. their one-point functions vanish). After doing so, the result for the two-point function takes a particularly compact form and exhibits a simple determinant structure:\footnote{A symmetric matrix of the type appearing here, which satisfies $M_{ij}=M_{i+j}$, is 
sometimes called ``persymmetric''. Similar persymmetric determinants often appear in integrable models, for instance in the correlation 
functions of the 2d Ising model \cite{perk1, perk2}. We thank J.H.H. Perk for pointing this out to us.} 
\beq
\langle \NO{\tilde\Phi^{L_1}}\NO{\tilde\Phi^{L_2}}\rangle =\frac{D_{L_1+1}}{D_{L_1}}\delta_{L_1,L_2}\comma \qquad \qquad D_L \equiv \det{}_{i,j}\left[\del_A^{i+j-2} \langle \mathcal{W}\rangle\right]\quad (1\leq i,j \leq L)\period 
\eeq
For higher-point functions, the result can be written succinctly in terms of certain polynomials $F_L(X)$, which by themselves are expressed in terms of determinants:
\beq
\langle \NO{\tilde\Phi^{L_1}}\NO{\tilde\Phi^{L_2}}\cdots \NO{\tilde\Phi^{L_m}}\rangle =\left.\left(\prod_{k=1}^{m}F_{L_k}(\del_{A^{\prime}})\right) \langle \mathcal{W}(A^{\prime})\rangle \right|_{A^{\prime}=A}\period
\eeq
See section \ref{sec:largeN} for further details including the definition of $F_L$.

As a special application of our analysis, we explain the relation of our correlators on the Wilson loop to the ``generalized Bremsstrahlung function'' $B_L(\theta)$ (whose definition is reviewed in more detail in section \ref{sec:Bremsstrahlung} below), which was computed previously in the planar large $N$ limit from integrability \cite{GS,GLS}. In particular we find that 
\beq
H_L(\theta)\equiv \frac{2\theta}{1-\frac{\theta^2}{\pi^2}}B_{L}(\theta) =\left.-\frac{1}{2}\del_{\theta}\log \frac{D_{L+1}}{D_L}\right|_{A=2\pi -2\theta}\period
\eeq
At large $N$, we show that this agrees with the integrability result. 

Our results are valid for the general $1/8$-BPS Wilson loop defined on an arbitrary contour on $S^2$, but perhaps the most interesting case is the $1/2$-BPS loop. Since the $1/2$-BPS loop is circular (or, by a conformal transformation, a straight line) it preserves a SL$(2,R)$ conformal subgroup, and therefore can be viewed as a conformal defect of the 4d theory. The correlation functions of operator insertions on the $1/2$-BPS loop are then constrained by the SL$(2,R)$ $d=1$ conformal symmetry, or more precisely by the 
$OSp(4^*|4) \supset {\rm SL}(2,R)\times SO(3)\times SO(5)$ 1d superconformal symmetry \cite{Drukker:2006xg}. Some of the properties of this defect CFT were recently studied at weak \cite{CDD,KK,KKKN,BGT} and strong coupling \cite{GRT,BGT}. The topological operators $\tilde \Phi^L$ correspond to a special kind of protected defect primaries $(Y\cdot \Phi)^L$, where $Y$ is a null polarization 5-vector. Such operators transform in the rank-$L$ symmetric traceless representation of $SO(5)$, and they belong to short representations of the 1d superconformal group, with protected scaling dimension $\Delta=L$. Because their 2-point and 3-point functions are fully fixed by the SL$(2,R)$ symmetry, the restriction to the topological choice of polarization vectors still allows one to extract exact results for the 2-point normalization and 3-point structure constants of the general defect primaries $(Y\cdot \Phi)^L$. Unlike the analogous case of single trace chiral primaries of the 4d theory, which are dual to protected closed string states, the structure constants in the 
present case are found to have a highly non-trivial dependence on the coupling constant. Our construction provides exact results for such structure constants of all operators in this protected subsector, which should provide valuable input for a conformal bootstrap approach to the Wilson-loop defect CFT (see e.g. \cite{LM}). 

The connection to integrability techniques emerges in the planar limit. At large $N$, we found that the results can be rewritten as a simple integral
\beq
\begin{aligned}
&\langle \NO{\tilde\Phi^{L_1}}\NO{\tilde\Phi^{L_2}}\cdots \NO{\tilde\Phi^{L_m}}\rangle = \oint d\mu \prod_{k=1}^{m}Q_{L_k}(x)\comma
\end{aligned}
\eeq
with the measure $d\mu$ given in \eqref{eq:defofmumeasure} (see also other forms of the measure \eqref{eq:symmeasure} and \eqref{eq:expmeasure}).
This is by itself an interesting result, but what is more intriguing is that the function $Q_L (x)$ that appears in the formula is directly related to the Quantum Spectral Curve \cite{QSC}, which is the most advanced method to compute the spectrum of the local operators in $\mathcal{N}=4$ SYM. The appearance of such functions 
in our setup hints at a potential applicability of the Quantum Spectral Curve to the problem of computing correlation functions.

The rest of this paper is organized as follows: In section \ref{sec:def}, we review the construction of the $1/8$ BPS Wilson loop and explain the definitions of the correlators that we study in this paper. We then relate them to the area derivative of the Wilson loop in section \ref{sec:result} by using the OPE and the Gram-Schmidt orthogonalization. The final result for the correlator at finite $N$ is given in subsection \ref{subsec:finalfiniteN}. In section \ref{sec:Bremsstrahlung}, we apply our method to compute the finite-$N$ generalization of the generalized Bremsstrahlung function. We then take the large $N$ limit of our results rewriting the correlators as an integral in section \ref{sec:largeN}. The results at large $N$ are expanded at weak and strong coupling in section \ref{sec:weakstrong} and compared against the direct perturbative computations on the gauge theory side and on the string theory side respectively. We also provide a matrix-model-like reformulation of the large $N$ results in section \ref{sec:matrixmodel}. Finally, section \ref{sec:conclusion} contains our conclusion and comments on future directions. 

\section{Topological correlators on the $1/8$ BPS Wilson Loop\label{sec:def}}
In this section, we explain the definitions of the topological correlators that we study in this paper and discuss their relation to the defect-CFT data.
\subsection{$1/8$ BPS Wilson loop}
\begin{figure}
\centering
\includegraphics[clip,height=4cm]{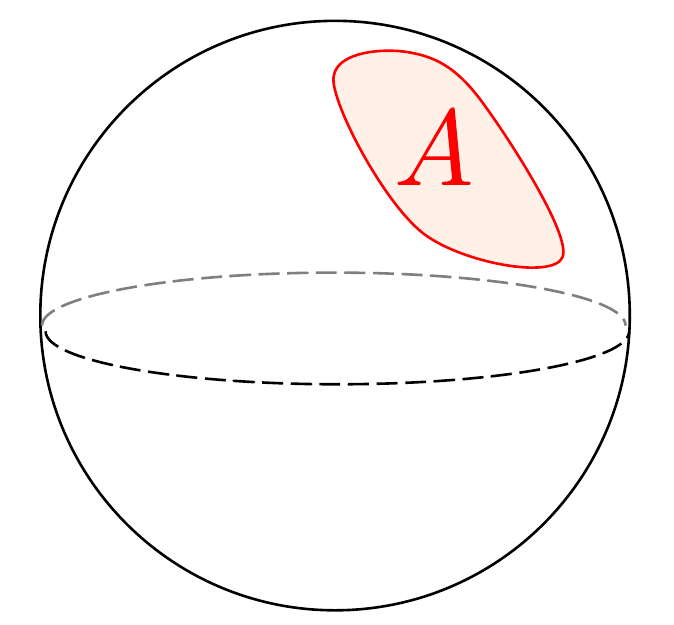}
\caption{General configuration of the $1/8$-BPS Wilson loop, denoted by a red curve. The $1/8$-BPS Wilson loop lives on $S^2$ and couples to a scalar as prescribed in \eqref{eq:defofWL}. The expectation value of such a loop depends only on the area $A$ of the region inside the loop (the red-shaded region in the figure). Note that, although ``the region inside/outside the loop'' is not a well-defined notion, such ambiguity does not affect the expectation value since it is invariant under $A\to 4\pi -A$, which exchanges the regions inside and outside the loop.}
\label{fig:fig0}
\end{figure}
The $1/8$ BPS Wilson loops is a generalization of the standard Wilson loop and it couples to a certain combination of the ${\cal N}=4$ SYM scalars, as well as the gauge field \cite{ESZ,DG,DrukkerQuarter,DGRT1,DGRT2}. In order to preserve $1/8$ of the superconformal symmetry, the contour $C$ must lie on a $S^2$ subspace of $R^4$, which we may take to be parametrized by $x_1^2+x_2^2+x_3^2=1$, and the coupling to the scalars is prescribed to be 
\beq\label{eq:defofWL}
\mathcal{W}\equiv \frac{1}{N} \tr \,{\rm P}\left[ e^{\oint_C \left(i A_j+ \epsilon_{kjl}x^{k}\Phi^{l}\right)dx^{j}}\right]\period
\eeq
where $i,j,k=1,2,3$, and we pick three out of the six scalar fields to be coupled to the loop. 
In what follows we will focus for simplicity on the fundamental Wilson loop, namely the trace in \eqref{eq:defofWL} is over the fundamental representation of the gauge group U$(N)$. However our construction below can be easily extended to arbitrary gauge group and arbitrary representations. 
  
The expectation value of this Wilson loop can be computed by supersymmetric localization\cite{Pestun1,Pestun2}. The result only depends on the rank of the gauge group $N$, the coupling constant $g_{\rm YM}$ and the area of the subregion inside the contour $C$, see figure \ref{fig:fig0}, which we denote by $A$ \cite{DrukkerQuarter,DGRT1,DGRT2,Pestun2}:
\beq
\left<\mathcal{W}\right>=\frac{1}{Z_{\rm MM}}\int [dM]\frac{1}{N}{\rm tr} \left(e^{M}\right) e^{-\frac{(4\pi)^2}{2 A (4\pi -A)g_{\rm YM}^2}\tr\left(M^2\right)}\period  
\eeq
To emphasize its dependence on the area, we sometimes denote $\left<\mathcal{W}\right>$ as $\left<\mathcal{W}(A)\right>$.
This matrix model integral can be evaluated explicitly \cite{DG} as 
\beq
\label{W-finite-N}
\left<\mathcal{W}\right>= \frac{1}{N}L^{1}_{N-1}\left(-\frac{\lambda^{\prime}}{4 N}\right) e^{\frac{\lambda^{\prime}}{8N}}\comma\qquad \lambda^{\prime}\equiv \lambda\left(1-\frac{a^2}{4\pi^2}\right)\comma
\eeq
with $\lambda$ being the 't Hooft coupling, $\lambda\equiv g_{\rm YM}^2 N$.
$L^{1}_{N-1}$ is the associated Laguerre polynomial and $a$ is defined by
\beq
a\equiv A-2\pi\period
\eeq
When $a=0$ ($A=2\pi$), the Wilson loop corresponds to a great circle of the $S^2$, and the operator (\ref{eq:defofWL}) couples to a single scalar field. This special case corresponds to the $1/2$-BPS Wilson loop, see section \ref{subsec:defectCFT} below. 

In the large $N$ limit, the result simplifies and can be expressed in terms of the Bessel function \cite{ESZ}:
\beq
\left.\left<\mathcal{W}\right>\right|_{\text{large N}}=\frac{2}{\sqrt{\lambda^{\prime}}}I_1 (\sqrt{\lambda^{\prime}})\period
\eeq
This can also be rewritten in terms of the $\theta$-deformed Bessel functions introduced in \cite{GLS},
\beq
I_n^{\theta}=\frac{1}{2}I_n \left(\sqrt{\lambda_{\theta} }\right)\left[\left(\frac{\pi+\theta}{\pi-\theta}\right)^{\frac{n}{2}}-(-1)^n\left(\frac{\pi-\theta}{\pi+\theta}\right)^{\frac{n}{2}}\right]\comma\qquad \lambda_{\theta}\equiv \lambda \left(1-\frac{\theta^2}{\pi^2}\right)\comma
\eeq
as
\beq\label{eq:WClargeN}
\left.\left<\mathcal{W}\right>\right|_{\text{large $N$}}=\frac{2}{\sqrt{\lambda}} I_1^{a/2}\period
\eeq
 
In \eqref{eq:WClargeN}, all the area dependence is encoded in the function $I_1^{a/2}$. This property turns out to be very useful when we later derive the integral expression for the topological correlators at large $N$.

\subsection{Correlators on the $1/8$ BPS loop}
The correlation function of the local operators on the Wilson loop is defined by\fn{Note that here we do not divide the correlator by the expectation value of the Wilson loop $\left< \mathcal{W}\right>$.}
\beq\label{eq:protectdef}
\langle O_1 (\tau_1)O_2(\tau_2)\cdots O_n(\tau_n)\rangle\equiv \left< \frac{1}{N}\tr \,{\rm P} \left[O_1(\tau_1)\cdots O_n(\tau_n)e^{\oint_C \left(i A_j+ \epsilon_{kjl}x^{k}\Phi^{l}\right)dx^{j}}\right]\right>_{\text{$\mathcal{N}=4$ SYM}}\period
\eeq
Here we parametrize the loop by $\tau\in [0,2\pi]$ and $\tau_i$'s are the positions of the operator insertions in that coordinate.

The BPS correlators we study in this paper are given by the following choice of the operators,
\beq
O_i (\tau_i)\equiv \tilde{\Phi}^{L_i}(\tau_i)\comma
\eeq 
where $\tilde{\Phi}$ is a position-dependent scalar\fn{Throughout this article, we use the convention in which the scalar propagator reads
\beq
\contraction{}{(\Phi_I}{(x_1))^{a}{}_{b}\quad(}{\Phi_J}
\left(\Phi_I(x_1)\right)^{a}{}_{b}\quad\left(\Phi_J(x_2)\right)^{c}{}_{d}=\frac{g_{\rm YM}^2\delta^{a}{}_{d}\delta^{c}{}_{b}\delta_{IJ}}{8\pi^2|x_1-x_2|^2}\comma
\eeq
where $a$-$d$ are the U$(N)$ matrix indices, and $I,J=1,\ldots,6$.
},
\beq\label{eq:posdepscalardef}
\tilde{\Phi}(\tau) =x_1 (\tau)\,\Phi_1+x_2 (\tau)\,\Phi_2+x_3 (\tau)\,\Phi_3+i\,\Phi_4\period
\eeq
An important property of such correlators is that they do not depend on the positions of the insertions $\tau_i$'s. This follows\fn{Alternatively, one should be able to show that the (twisted) translation generator which moves the positions of the insertions is $\mathcal{Q}$-exact with $\mathcal{Q}$ being one of the supercharges preserved by the configuration. It then follows that the correlators are position-independent. In the absence of Wilson loops, this was shown in \cite{DP}, which studied correlation 
functions of operators precisely of this kind. See also \cite{CLPY} for a similar discussion in the CFT$_3$ context.} from the fact that, after localization, these operators are mapped to field-strength insertions in two-dimensional Yang-Mills theory (see section \ref{subsec:fromOPE} for further explanation), which enjoys invariance under area-preserving diffeomorphisms, making it almost topological. Because of their position independence, we will call them ``topological correlators'' in the rest of this paper.

\subsection{$1/2$-BPS Wilson loop and defect CFT}\label{subsec:defectCFT}
For special contours, the Wilson loops preserve higher amount of supersymmetry. Particularly interesting 
among them is the $1/2$-BPS Wilson loop,  whose contour is a circle along the equator and which couples to a single scalar $\Phi_3$ \cite{ESZ,DG}:
\beq
\mathcal{W}_{\text{1/2-BPS}}\equiv \frac{1}{N}\tr \,{\rm P}\,\exp\left(\oint_{\rm equator} \left(i A_j\dot{x}^j+ \Phi_{3}|\dot{x}|\right)d\tau\right)
\eeq
Since the contour is circular, the $1/2$-BPS loop preserves the SL$(2,R)$ conformal symmetry. Therefore, one can view the correlators on the $1/2$-BPS Wilson loop as correlators of a defect CFT.
To make this point precise, one needs to consider the {\it normalized} correlator, which is obtained by dividing the bare correlator \eqref{eq:protectdef} by the expectation value of the Wilson loop:
\beq
\langle \!\langle O_1 (\tau_1) O_2 (\tau_2)\cdots O_{n}(\tau_n)\rangle \!\rangle \equiv \frac{\langle O_1 (\tau_1) O_2 (\tau_2)\cdots O_{n}(\tau_n)\rangle}{\langle \mathcal{W}\rangle}\period
\eeq
After the normalization, the expectation value of the identity operator becomes unity and the correlators obey the standard properties of the defect CFT correlators.

Using these normalized correlators, one can extract the defect CFT data from the topological correlators. To see this, let us consider general two- and three-point functions of BPS operators on the the $1/2$-BPS loop:
\beq\label{eq:generalBPS}
\begin{aligned}
G_{L_1,L_2}&=\langle\!\langle (Y_1\cdot \vec{\Phi})^{L_1}(\tau_1)\,\,\, (Y_2\cdot \vec{\Phi})^{L_2}(\tau_2)\rangle\!\rangle_{\rm circle}\comma\\
G_{L_1,L_2,L_3}&= \langle\!\langle (Y_1\cdot \vec{\Phi})^{L_1}(\tau_1)\,\,\, (Y_2\cdot \vec{\Phi})^{L_2}(\tau_2)\,\,\,(Y_3\cdot \vec{\Phi})^{L_3}(\tau_3)\rangle\!\rangle_{\rm circle}\period
\end{aligned}
\eeq
In \eqref{eq:generalBPS}, $\vec{\Phi}\equiv (\Phi_1,\Phi_2,\Phi_4,\Phi_5,\Phi_6)$ and $Y_i$'s are five-dimensional complex vectors satisfying $Y_i\cdot Y_i=0$.
Unlike the topological correlators \eqref{eq:protectdef}, the correlators \eqref{eq:generalBPS} depend on the positions of the insertions, and the vectors $Y_i$. However, because of the conformal symmetry and the SO$(5)$ R-symmetry, their dependence is completely fixed to be\fn{Of course, one may also write the analogous result for the straight line geometry, which is related to the circle by a conformal transformation}:
\beq\label{eq:depgeneralBPS}
\begin{aligned}
G_{L_1,L_2}&=n_{L_1}(\lambda,N)\times \frac{\delta_{L_1,L_2}(Y_1\cdot Y_2)^{L_1}}{(2\sin \frac{\tau_{12}}{2})^{2L_1}}\comma\\
G_{L_1,L_2,L_3}&=c_{L_1,L_2,L_3}(\lambda,N)\times \frac{(Y_1\cdot Y_2)^{L_{12|3}}(Y_2\cdot Y_3)^{L_{23|1}}(Y_3\cdot Y_1)^{L_{31|2}}}{\left(2\sin \frac{\tau_{12}}{2}\right)^{2 L_{12|3}}\left(2\sin \frac{\tau_{23}}{2}\right)^{ 2L_{23|1}}\left(2\sin \frac{\tau_{31}}{2}\right)^{2 L_{31|2}}}\comma
\end{aligned}
\eeq
with $\tau_{ij}\equiv \tau_i-\tau_j$ and $L_{ij|k}\equiv (L_i+L_j-L_k)/2$. Here $n_{L_1}$ is the normalization of the two-point function while $c_{L_1,L_2,L_3}$ is the structure constant. As shown above, both of these quantities are nontrivial functions of $\lambda$ and $N$. Note that, although we often set the normalization of the two-point function to be unity in conformal field theories, for special operators the normalization itself can have physical meaning\fn{Other examples are the stress-energy tensor and the conserved currents, whose two-point functions are related to $C_T$ and $C_J$.}. For instance, the length-$1$ operator $(Y\cdot \vec{\Phi})$ is related to the displacement operator and has a canonical normalization which is related to the Bremsstrahlung function \cite{CHMS}.

Now, if we go to the topological configuration by setting the vectors $Y_i$ to be 
\beq\label{eq:choiceY}
Y_i =(\cos \tau_i,\sin \tau_i,0,i,0,0) \comma
\eeq
we get
\beq
\begin{aligned}
\left. G_{L_1,L_2}\right|_{\rm topological}&=\left(-\frac{1}{2}\right)^{L_1}\times n_{L_1}\delta_{L_1,L_2}\comma\\
\left.G_{L_1,L_2,L_3}\right|_{\rm topological}&=\left(-\frac{1}{2}\right)^{\frac{L_1+L_2+L_3}{2}}\times c_{L_1,L_2,L_3}\period
\end{aligned}
\eeq
This shows that the topological correlators compute the normalization and the structure constant in the defect CFT up to trivial overall factors. Alternatively, one can consider the ratio
\beq\label{eq:standarddefofc}
\left.\frac{G_{L_1,L_2,L_3}}{(G_{L_1,L_1}G_{L_2,L_2}G_{L_3,L_3})^{1/2}}\right|_{\rm topological}=\frac{c_{L_1,L_2,L_3}}{(n_{L_1}n_{L_2}n_{L_3})^{1/2}}\comma
\eeq
and get rid of the overall factors. The quantity which appears on the right hand side of \eqref{eq:standarddefofc} is a structure constant in the standard CFT normalization; namely the normalization in which the two-point function becomes unity.

Note that, for higher-point functions, there is no such a direct relation between the general correlators and the topological correlators: The general higher-point correlators are nontrivial functions of the cross ratios while the topological correlators do not depend at all on the positions. Thus for higher-point functions, one cannot reconstruct the general correlators just from the topological correlators. 

\section{Computation of the correlators\label{sec:result}}
We now compute the correlators on the $1/8$ BPS Wilson loop
\beq\label{eq:defdefcorre}
 \langle \tilde{\Phi}^{L_1}(\tau_1)\tilde{\Phi}^{L_2}(\tau_2)\cdots \tilde{\Phi}^{L_n}(\tau_n)\rangle \comma
\eeq
using the results from localization. We first discuss the correlators on the $1/2$ BPS Wilson loop from the OPE perspective and then present a general method that applies also to the $1/8$ BPS Wilson loop.
\subsection{Correlators on the $1/2$ BPS Wilson loop from OPE}\label{subsec:fromOPE}
When all the operators are length-1  and the Wilson loop is circular (or equivalently $1/2$ BPS), the correlators \eqref{eq:defdefcorre} were already computed in \cite{GRT}. Let us fist briefly review their computation: By performing localization, one can reduce the computation of the $1/8$ BPS Wilson loop in $\mathcal{N}=4$ SYM to the computation of the Wilson loop in two-dimensional Yang-Mills theory in the zero instanton sector \cite{Pestun2}. Under this reduction, the insertion of the position-dependent scalar $\tilde{\Phi}$ is mapped to the insertion of the dual field strength $\ast F_{2d}$:
\beq
\tilde{\Phi}\quad  \Leftrightarrow \quad i\ast F_{2d}\period
\eeq
 Using this correspondence\fn{At weak coupling, this correspondence was checked by the direct perturbative computation on both sides in \cite{BGPS}.}, one can insert $\tilde{\Phi}$'s on the circular Wilson loop by differentiating its expectation value with respect to the area $A$:
 \beq\label{eq:1/2BPSsingle}
\langle\underbracket{\tilde{\Phi}\cdots \tilde{\Phi}}_{L}\rangle|_{\text{circle}}= \left.\frac{\del^L\!\left< \mathcal{W}\right>}{(\del A)^L}\right|_{A=2\pi}\period
 \eeq
Using the expression \eqref{eq:1/2BPSsingle}, one can compute arbitrary correlation functions of single-letter insertions $\tilde{\Phi}$.

To study more general BPS correlators, we also need to know how to insert operators of longer length, $\tilde{\Phi}^{L}$ with $L>1$.  The first guess might be to relate it simply to the $L$-th derivative of the Wilson loop,
\beq
\tilde{\Phi}^{L}\overset{\huge ?}{\sim} \frac{\del^L\left< \mathcal{W}[C]\right>}{(\del A)^L}\period
\eeq
This guess, however, turns out to be incorrect. To see why it is so, let us consider $\tilde{\Phi}^2$ as an example. We know that the second derivative of $\langle \mathcal{W}\rangle$ corresponds to the insertion of two $\tilde{\Phi}$'s on the Wilson loop. Since the correlator we are studying is topological, one can bring the two $\tilde{\Phi}$'s close to each other without affecting the expectation value and rewrite them using the operator product expansion. This procedure does produce the length-2 operator $\tilde{\Phi}^2$ as we wanted, but the problem is that it also produces other operators\fn{Owing to the representation theory of SO(5), the OPE does not produce higher-charge operators, $\tilde{\Phi}^{k}$ with $k>2$.}:
\beq\label{eq:topologicalOPE}
\overbracket{\tilde{\Phi}\quad  \tilde{\Phi}}^{\rm OPE} = \tilde{\Phi}^2 + c_1 \, \tilde{\Phi} + c_0\, {\bf 1} \period
\eeq 
Here $c_i$'s are some numerical coefficients and ${\bf 1}$ is the identity operator.
Thus, to really get the length-2 operator, one has to subtract these unnecessary OPE terms from $\tilde{\Phi} \tilde{\Phi}$:
\beq\label{eq:defoflength2}
\NO{\tilde{\Phi}^2} = \tilde{\Phi}\tilde{\Phi} -c_1 \tilde{\Phi}-c_0{\bf 1}\period
\eeq
Here $\tilde{\Phi}\tilde{\Phi}$ on the right hand side denotes two single-letter insertions at separate points while $\NO{\tilde{\Phi}^2}$ is a length-2 operator inserted at a single point.  Since this subtraction procedure is conceptually similar to the normal ordering, we hereafter put the normal-ordering symbol $\NO{\ast}$ to the operator obtained in this way.

The coefficients $c_i$'s are nothing but the OPE coefficients of the topological OPE \eqref{eq:topologicalOPE}. They are thus related to the following three-point functions:
\beq\label{eq:c1propto}
c_1 \propto  \langle \tilde{\Phi}\tilde{\Phi}\tilde{\Phi}\rangle|_{\rm circle} \comma \qquad c_0\propto  \langle \tilde{\Phi}\tilde{\Phi}{\bf 1}\rangle|_{\rm circle}\period
\eeq
If we were using the operators, $\tilde{\Phi}/\langle \tilde{\Phi}\tilde{\Phi}\rangle^{1/2}$ whose two-point function is unit-normalized, the constants of proportionality in \eqref{eq:c1propto} would have been unity. However the operators we are using here are not unit-normalized and one has to take into account that effect. This leads to the following expressions for the coefficients $c_1$ and $c_0$:
\beq
\begin{aligned}
&c_1 = \frac{\langle  \tilde{\Phi}\tilde{\Phi}\tilde{\Phi}\rangle|_{\rm circle}}{ \langle \tilde{\Phi}\tilde{\Phi}\rangle|_{\rm circle}} =\left. \frac{ \del_A^{3}\left< \mathcal{W}\right>}{\del_A^{2}\left< \mathcal{W}\right>}\right|_{A=2\pi}\comma\\
&c_0=\frac{ \langle \tilde{\Phi}\tilde{\Phi}{\bf 1}\rangle|_{\rm circle}}{ \langle {\bf 1}{\bf 1}\rangle|_{\rm circle}}= \left. \frac{\del_A^{2}\left< \mathcal{W}\right>}{\left< \mathcal{W}\right>}\right|_{A=2\pi}\period
\end{aligned}
\eeq

We can repeat this procedure to express operators of arbitrary length in terms of single-letter insertions and compute their correlation functions. Although these procedures can be easily automated using computer programs, they do not give much insight into the underlying structure. In the next section, we discuss a simpler way to reorganize these procedures which also leads to a simple closed-form expression.

\subsection{Construction of operators from the Gram-Schmidt orthogonalization\label{subsec:constructionofoperatorfrom}}
As a direct consequence of the subtraction procedures \eqref{eq:defoflength2}, the operators constructed above satisfy the following important properties:
\begin{itemize}
\item $
\langle  \NO{\tilde{\Phi}^{L}}\NO{\tilde{\Phi}^{M}}\rangle|_{\rm circle}\propto \delta_{LM}\period$
\item $\NO{\tilde{\Phi}^{L}}$ is a linear combination of $\underbracket{\tilde{\Phi}\cdots\tilde{\Phi}}_{M}$ with $M\leq L$.
\item The coefficient of the $M=L$ term is $1$. Namely $\NO{\tilde{\Phi}^{L}}=\underbracket{\tilde{\Phi}\cdots\tilde{\Phi}}_{L}+\cdots$.
\end{itemize}
The operator basis with such properties turns out to be unique and can be constructed systematically by using the so-called {\it Gram-Schmidt orthogonalization}. As we see below, it also allows us to write down a closed-form expression for the operators $\NO{\tilde{\Phi}^{L}}$. 

The Gram-Schmidt orthogonalization is an algorithmic way of getting the orthogonal basis from a given set of vectors. It was recently applied in the computation of Coulomb branch operators in $\mathcal{N}=2$ superconformal theories in \cite{GGIKKP}. Its large $N$ limit was discussed in \cite{BNPV} while the case for $\mathcal{N}=4$ SYM was analyzed further in \cite{Gomez1, Gomez2}. What we describe below is a new application of the method to the correlators on the Wilson loop. To get a glimpse of how it works, let us orthogonalize two arbitrary vectors $\{{\bf v}_1\comma {\bf v}_2\}$. A simple way of doing so is to define new vectors as
\beq\label{eq:GS1}
{\bf u}_1 = {\bf v}_1 \comma \qquad {\bf u}_2 = {\bf v_2}-\frac{\langle{\bf v}_1,{\bf v}_2 \rangle }{\langle{\bf v}_1,{\bf v}_1 \rangle}{\bf v_1}\comma
\eeq
where $\langle\ast ,\ast \rangle$ denotes the inner product between two vectors. This is of course just an elementary manipulation, but the key point is that one can re-express \eqref{eq:GS1} as
\beq
{\bf u}_1 ={\bf v}_1 \comma \qquad {\bf u}_2 =\frac{1}{\langle{\bf v}_1,{\bf v}_1 \rangle}  \dmatrix{cc}{\langle{\bf v}_1,{\bf v}_1 \rangle&\langle{\bf v}_1,{\bf v}_2 \rangle\\{\bf v}_1&{\bf v}_2}\comma
\eeq
where $|\ast |$ denotes a determinant of a matrix.
This expression can be readily generalized to the case with more vectors. The result reads
\beq
\begin{aligned}
&{\bf u}_k =\frac{1}{d_{k-1}}\dmatrix{cccc}{\langle{\bf v}_1,{\bf v}_1 \rangle&\langle{\bf v}_1,{\bf v}_2 \rangle&\cdots&\langle{\bf v}_1,{\bf v}_k \rangle\\\langle{\bf v}_2,{\bf v}_1 \rangle&\langle{\bf v}_2,{\bf v}_2 \rangle&\cdots&\langle{\bf v}_2,{\bf v}_k \rangle\\\vdots&\vdots&\ddots&\vdots\\\langle{\bf v}_{k-1},{\bf v}_1 \rangle&\langle{\bf v}_{k-1},{\bf v}_2 \rangle&\cdots&\langle{\bf v}_{k-1},{\bf v}_k \rangle\\{\bf v}_1&{\bf v}_2&\cdots&{\bf v}_k}\comma\\
&d_k= \dmatrix{cccc}{\langle{\bf v}_1,{\bf v}_1 \rangle&\langle{\bf v}_1,{\bf v}_2 \rangle&\cdots&\langle{\bf v}_1,{\bf v}_k \rangle\\\langle{\bf v}_2,{\bf v}_1 \rangle&\langle{\bf v}_2,{\bf v}_2 \rangle&\cdots&\langle{\bf v}_2,{\bf v}_k \rangle\\\vdots&\vdots&\ddots&\vdots\\\langle{\bf v}_{k},{\bf v}_1 \rangle&\langle{\bf v}_{k},{\bf v}_2 \rangle&\cdots&\langle{\bf v}_{k},{\bf v}_k \rangle}\period
\end{aligned}
\eeq
For details of the derivation, see standard textbooks on linear algebra. The new vectors defined above are orthogonal but not normalized. Their norms can be computed using the definitions above and we get
\beq\label{eq:GSdet}
\langle {\bf u}_k\comma {\bf u}_l\rangle =\frac{d_k}{d_{k-1}}\delta_{kl}\period
\eeq

We now apply the Gram-Schmidt orthogonalization to the set of single-letter insertions $\{{\bf 1}\,,\, \tilde{\Phi}\,,\, \tilde{\Phi}\tilde{\Phi},\ldots\}$.  The norms between these vectors are given by the two-point functions, which can be computed by taking derivatives of $\left< \mathcal{W}\right> $,
\beq
\langle \underbracket{\tilde{\Phi}\cdots \tilde{\Phi}}_{L}\,\,\underbracket{\tilde{\Phi}\cdots \tilde{\Phi}}_{M}\rangle = (\del_A)^{L+M}\left< \mathcal{W}\right>
\eeq
We then get the expression for the operator $\NO{\tilde{\Phi}^{L}}$,
\beq
\begin{aligned}\label{eq:halfphiD}
\NO{\tilde{\Phi}^{L}}&=\frac{1}{D_{L}}\dmatrix{cccl}{\left<\mathcal{W}\right>&\left<\mathcal{W}\right>^{(1)}&\cdots&\left<\mathcal{W}\right>^{(L)}\\\left<\mathcal{W}\right>^{(1)}&\left<\mathcal{W}\right>^{(2)}&\cdots&\left<\mathcal{W}\right>^{(L+1)}\\ \vdots&\vdots&\ddots&\quad\vdots\\\left<\mathcal{W}\right>^{(L-1)}&\left<\mathcal{W}\right>^{(L)}&\cdots&\left<\mathcal{W}\right>^{(2L-1)} \\{\bf 1}&\tilde{\Phi}&\cdots&\underbracket{\tilde{\Phi}\cdots\tilde{\Phi}}_L}\comma\\
D_L&= \dmatrix{cccc}{\left<\mathcal{W}\right>&\left<\mathcal{W}\right>^{(1)}&\cdots&\left<\mathcal{W}\right>^{(L-1)}\\\left<\mathcal{W}\right>^{(1)}&\left<\mathcal{W}\right>^{(2)}&\cdots&\left<\mathcal{W}\right>^{(L)}\\\vdots&\vdots&\ddots&\vdots\\\left<\mathcal{W}\right>^{(L-1)}&\left<\mathcal{W}\right>^{(L)}&\cdots&\left<\mathcal{W}\right>^{(2L-2)}}\comma
\end{aligned}
\eeq
with $\left< \mathcal{W}\right>^{(k)} \equiv (\del_A)^{k}\left< \mathcal{W}\right>$. Let us emphasize that this method applies to general $1/8$ BPS Wilson loops. To get the result for the $1/2$ BPS loop, one just needs to set $A=2\pi$ at the end of the computation. 
For small values of $n$, one can check explicitly that this expression coincides with the operators obtained by the recursive procedure outlined in the previous subsection.  One can also check that the basis obtained in this way satisfies the aforementioned three properties. 

Here we assumed that all the operators that appear in the OPE are multi-letter insertions on the Wilson loop, namely $\tilde{\Phi}^{L}$ inserted on $\mathcal{W}$. This is indeed true at the planar limit. However, at the non-planar level, the operator product expansion can also produce ``multi-trace operators'', for instance $\mathcal{W}\,\,{\rm tr}[\tilde{\Phi}^{L}]$. In the presence of such extra operators, one has to appropriately modify the Gram-Schmidt procedure. The details of the multi-trace operators and their effect on the Gram-Schmidt process are explained in our follow-up paper \cite{UpComing}. Since we assume that such operators are absent, for generic $L$ the analysis in this paper applies only to the large $N_c$ limit. There are however a few exceptions where the multi-trace operators do not show up even at the non-planar level. One example is the operator with $L=1$, namely $\tilde{\Phi}$. Since there is only one operator with $L<1$, which is the identity operator, the multi-trace operators cannot affect the orthogonalization of this operator. Another example is $\tilde{\Phi}^2$ for the 1/2-BPS Wilson loop. For the $1/2$-BPS Wilson loop, the insertion $\tilde{\Phi}$ cannot get contracted against the Wilson loop. Therefore, the OPE of $\tilde{\Phi}\tilde{\Phi}$ cannot produce the operator with charge $1$, such as $\tilde{\Phi}$ or ${\rm tr}[\tilde{\Phi}]$. Therefore, the orthogonalization of $\tilde{\Phi}^2$ only involves the identity operator and is not affected by the multi-trace operators\fn{Yet another example is the operator $\tilde{\Phi}^3$ defined on the $1/2$ BPS Wilson loop in the {\it SU$(N_c)$} SYM. This is because the multi-trace operator that can potentially mix with $\tilde{\Phi}^3$, ${\rm tr}[\tilde{\Phi}]$, is absent in the SU$(N_c)$ SYM.}. It is also worth mentioning that the construction in this paper works in the opposite limit, namely the U$(1)$ SYM. In that case, there is only one operator for each $R$-charge and the mixing problem discussed above is absent.

Owing to the property \eqref{eq:GSdet}, the two-point function of the operators $\tilde{\Phi}^{L}$ is given by a ratio of determinants:
\beq
\langle \NO{\tilde{\Phi}^{L}}\NO{\tilde{\Phi}^{M}}\rangle=\frac{D_{L+1}}{D_{L}}\delta_{LM}\period
\eeq
For the $1/2$ BPS loop, this provides an exact result for the normalization of the two-point function in the defect CFT (see the discussions in section \ref{subsec:defectCFT}),
\beq
n_L =\left.(-2)^{L}\frac{D_{L+1}}{D_{L}}\right|_{A=2\pi}\period
\eeq 
As it is well-known, the result for $L=1$ is related to the normalization of the displacement operators while the results for $L>0$ provide new defect-CFT observables\fn{Although the normalization of the operators is usually not meaningful, for this class of operators, there is a canonical normalization induced by the facts that $\tilde{\Phi}$ is related to the displacement operator and $\NO{\tilde{\Phi}^{L}}$ is essentially a product of $L$ $\tilde{\Phi}$'s.}. 

We will later see in section \ref{subsec:comparison} that the large-$N$ limit of these determinants is related to the determinant representation of the generalized Bremsstrahlung function derived previously in \cite{GS,GLS}.
\subsection{A remark on the $1/8$ BPS Wilson loop}
As mentioned above, the Gram-Schmidt process can be applied to the general $1/8$ BPS Wilson loops. At the level of formulas, one just needs to keep the area $A$ general in \eqref{eq:halfphiD} and \eqref{eq:generalhpt}. However, there is one important qualitative difference which we explain below.

Unlike the $1/2$ BPS Wilson loop, the first-order derivative $\left< \mathcal{W}\right>^{(1)}$ does not vanish for the general $1/8$ BPS Wilson loop. This means that the single-letter insertion $\tilde{\Phi}$ has a non-vanishing one-point function; in other words, the two-point function of $\tilde{\Phi}$ and the identity operator ${\bf 1}$ is nonzero. Therefore, to define an orthogonal set of operators, one has to perform the subtraction even for $\tilde{\Phi}$. In fact, by applying the Gram-Schmidt orthogonalization, we get
\beq
\NO{\tilde{\Phi}}=\tilde{\Phi}-\frac{\left<\mathcal{W}\right>^{(1)}}{\left<\mathcal{W}\right>}{\bf 1}\period
\eeq
We thus need to distinguish $\NO{\tilde{\Phi}}$ from $\tilde{\Phi}$. This was one of the reasons why we preferred to put the normal-ordering symbol when defining the operator $\NO{\tilde{\Phi}^{J}}$.
\subsection{Results for topological correlators\label{subsec:finalfiniteN}}
Using the closed-form expression \eqref{eq:halfphiD}, one can compute higher-point functions of $\NO{\tilde{\Phi}^{N}}$. To express the result, it is convenient to introduce a polynomial
\beq\label{eq:FnXexplicit}
F_L(X)=\frac{1}{D_{L}}\dmatrix{cccl}{\left<\mathcal{W}\right>&\left<\mathcal{W}\right>^{(1)}&\cdots&\left<\mathcal{W}\right>^{(L)}\\\left<\mathcal{W}\right>^{(1)}&\left<\mathcal{W}\right>^{(2)}&\cdots&\left<\mathcal{W}\right>^{(L+1)}\\ \vdots&\vdots&\ddots&\quad\vdots\\\left<\mathcal{W}\right>^{(L-1)}&\left<\mathcal{W}\right>^{(L)}&\cdots&\left<\mathcal{W}\right>^{(2L-1)} \\1&X&\cdots&X^L}\period
\eeq
By replacing $X^k$ by $\underbracket{\tilde{\Phi}\cdots\tilde{\Phi}}_k$, one recovers $\NO{\tilde{\Phi}^{L}}$. In terms of these polynomials, the higher-point function reads
\beq\label{eq:generalhpt}
\begin{aligned}
\langle  \NO{\tilde{\Phi}^{L_1}}\NO{\tilde{\Phi}^{L_2}}\cdots \NO{\tilde{\Phi}^{L_m}}\rangle =\left(\prod_{k=1}^{m}F_{L_k} (\del_{A^{\prime}})\right) \left<\mathcal{W}(A^{\prime})\right>|_{A^{\prime}=A}\period
\end{aligned}
\eeq
Let us make two remarks regarding this formula: First, the derivatives $\del_{A^{\prime}}$'s on the right hand side act only on the last term $\left<\mathcal{W}(A^{\prime})\right>$ (not on the coefficients of the polynomials $F_{L_k}$). 
Second, the polynomial $F_L$ is not just a technical tool for writing down higher-point correlators, but it gives an explicit map between the OPE and the multiplication of polynomials. To see this, consider a product of two such polynomials. Since the product is also a polynomial, one can express it as a sum of $F_L$'s,
\beq
F_{L_1}(X)F_{L_2}(X)=\sum_{M=0}^{L_1+L_2}\bar{c}_{L_1,L_2,M}F_{M}(X)\comma
\eeq
where $\bar{c}_{L_1,L_2,M}$ is a ``structure constant'' for the multiplication of polynomials. This expansion can be performed also on the right hand side of  \eqref{eq:generalhpt}. On the other hand,  we can perform a similar expansion on the left hand side of \eqref{eq:generalhpt} using the OPE,
\beq
\NO{\tilde{\Phi}^{L_1}}\NO{\tilde{\Phi}^{L_2}}=\sum_{M=1}^{L_1+L_2}c_{L_1,L_2,M}\NO{\tilde{\Phi}^{M}}\period
\eeq
Equating the two expressions, we conclude that these two structure constants must coincide, namely $\bar{c}_{L_1,L_2,M}=c_{L_1,L_2,M}$. This provides an interesting correspondence between the multiplication of polynomials and the OPE. 

We can also express the results more explicitly in terms of determinants. For this purpose, we first perform the Laplace expansion of the polynomial $F_L (X)$:
\beq
F_L(X)=\frac{1}{D_L} \sum_{n=0}^{L}(-1)^{L+n}D_{L+1}^{(L+1,n+1)}X^{n}\period
\eeq
Here $D_L^{(i,j)}$ is a minor of $D_L$ obtained by deleting the $i$-th row and $j$-th column. 
We then substitute this expression into \eqref{eq:generalhpt} to get
\beq\label{eq:expdetcorrelator}
\begin{aligned}
\langle  \NO{\tilde{\Phi}^{L_1}}\NO{\tilde{\Phi}^{L_2}}\cdots \NO{\tilde{\Phi}^{L_m}}\rangle =\sum_{n_1=0}^{L_1}\cdots\sum_{n_m=0}^{L_m}\left(\prod_{k=1}^{m}(-1)^{L_k+n_k}\frac{D_{L_k+1}^{(L_k+1,n_k+1)}}{D_{L_k}}\right)\langle \mathcal{W}\rangle^{(n_{\rm tot})}\comma
\end{aligned}
\eeq
with $n_{\rm tot} \equiv \sum_{k=1}^{m}n_k$. 

We can also perform one of the sums explicitly to reconstruct a determinant: The result reads
\beq
\begin{aligned}\label{eq:expdetcorrelator2}
\langle  \NO{\tilde{\Phi}^{L_1}}\NO{\tilde{\Phi}^{L_2}}\cdots \NO{\tilde{\Phi}^{L_m}}\rangle =\sum_{n_2=0}^{L_2}\cdots\sum_{n_m=0}^{L_m}\left(\prod_{k=2}^{m}(-1)^{L_k+n_k}\frac{D_{L_k+1}^{(L_k+1,n_k+1)}}{D_{L_k}}\right)\tilde{D}_{L_1,n^{\prime}_{\rm tot}}\comma
\end{aligned}
\eeq
where $n^{\prime}_{\rm tot}=\sum_{k=2}^{m}n_k$ and $\tilde{D}_{L,n}$ is given by
\beq
\begin{aligned}\label{eq:tildeDLnp}
\tilde{D}_{L,n}&\equiv  \dmatrix{cccc}{\left<\mathcal{W}\right>&\left<\mathcal{W}\right>^{(1)}&\cdots&\left<\mathcal{W}\right>^{(L)}\\\left<\mathcal{W}\right>^{(1)}&\left<\mathcal{W}\right>^{(2)}&\cdots&\left<\mathcal{W}\right>^{(L+1)}\\\vdots&\vdots&\ddots&\vdots\\\left<\mathcal{W}\right>^{(L-1)}&\left<\mathcal{W}\right>^{(L)}&\cdots&\left<\mathcal{W}\right>^{(2L-1)}\\\left<\mathcal{W}\right>^{(n)}&\left<\mathcal{W}\right>^{(n+1)}&\cdots&\left<\mathcal{W}\right>^{(L+n)}}\period
\end{aligned}
\eeq
Importantly, $\tilde{D}_{L,n}$ vanishes unless $n\geq L$ since otherwise the last row coincides with one of the rows above. This allows us to restrict the sum in \eqref{eq:expdetcorrelator2} to $n_{\rm tot}^{\prime}\geq L_1$. In particular, for ``extremal'' correlators which satisfy $L_1=\sum_{k=2}^{m}L_k$, there is only one term in the sum that survives and we get a simpler formula 
\beq
\begin{aligned}\label{eq:expdetcorrelator3}
\langle  \NO{\tilde{\Phi}^{L_1}}\NO{\tilde{\Phi}^{L_2}}\cdots \NO{\tilde{\Phi}^{L_m}}\rangle =\frac{D_{L_1+1}}{D_{L_1}}\qquad \text{for }L_1=\sum_{k=2}^{m}L_k\period
\end{aligned}
\eeq

For general correlators, the expression \eqref{eq:expdetcorrelator2} is not very concise as it involves several terms. The results for two- and three-point functions of operators with $L\leq 3$ are given explicitly in Appendix \ref{ap:explicit}. We will later see in sections \ref{sec:largeN} and \ref{sec:matrixmodel} that in the large $N$ limit there is an elegant reformulation in terms of integrals and a matrix model.
\section{Generalized Bremsstrahlung functions\label{sec:Bremsstrahlung}}
As an application of our method, in this section we compute the so-called ``generalized Bremsstrahlung function''. The result provides finite-$N$ generalization of the planar results computed previously in \cite{GS,GLS} using integrability \cite{DrukkerTBA,CMS}.

\subsection{Cusp anomalous dimension and Bremsstrahlung function}
\begin{figure}
\centering
\includegraphics[clip,height=3cm]{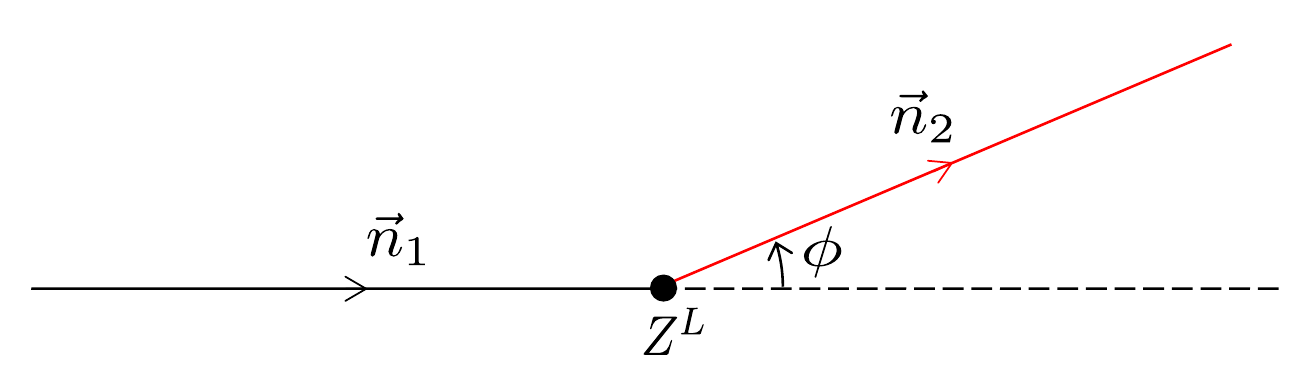}
\caption{Cusped Wilson line with insertions. The cusped Wilson line consists of two semi-infinite lines which intersects with an angle $\phi$ at the origin, and the insertions $Z^{L}$. The scalar coupling of each semi-infinite line is given by the vector $\vec{n}_{1,2}$, and the relative angle between the two vectors is $\theta$. The divergence from this Wilson line is controlled by the generalized Bremsstrahlung function.}
\label{fig:fig1}
\end{figure}
Let us first recall the definition of the generalized Bremsstrahlung function. Consider the following cusped Wilson line with insertions (see also figure \ref{fig:fig1}):
\beq
\mathcal{W}_L(\theta,\phi) \equiv {\rm P} \exp \int^{0}_{-\infty}d\tau\left[i A\cdot \dot{x}_1+\vec{\Phi}\cdot \vec{n}_1|\dot{x}_1|\right]\times Z^{L}\times {\rm P} \exp \int^{\infty}_{0}d\tau\left[i A\cdot \dot{x}_2+\vec{\Phi}\cdot \vec{n}_2|\dot{x}_2|\right]\period
\eeq
Here $Z =\Phi_3+i\Phi_4$ and the $x_{1,2}(t)$ and $\vec{n}_{1,2}$ are given by
\beq
\begin{aligned}
&\dot{x}_1= (1,0,0,0)\comma\qquad &&\dot{x}_2=(\cos\phi,\sin\phi,0,0)\comma\\
&n_1=(1,0,0,0,0)\comma\qquad &&n_2=(\cos \theta,\sin\theta,0,0,0,0)\period
\label{cusped-R2}
\end{aligned}
\eeq
As shown above, $\mathcal{W}_L$ is parametrized by the two angles $\theta$ and $\phi$. When $\theta=\phi$, $\mathcal{W}_L$ is BPS and the expectation value $\langle \mathcal{W}_L\rangle$ is finite. However, if $\theta\neq \phi$, it has the divergence controlled by the cusp anomalous dimension $\Gamma_L$:
\beq\label{eq:cuspanodef}
\langle \mathcal{W}_L(\theta,\phi)\rangle \sim \left(\frac{\epsilon_{\rm UV}}{r_{\rm IR}}\right)^{\Gamma_{L}(\theta,\phi)}
\eeq
Here $\epsilon_{\rm UV}$ and $r_{\rm IR}$ are the UV and IR (length) cutoffs respectively.

The cusp anomalous dimension can be expanded near $\theta \sim \phi$ and the leading term in the expansion reads
\beq
\begin{aligned}
&\Gamma_{L}(\theta,\phi)=(\theta-\phi)H_L (\theta)+O((\theta-\phi)^2)\comma
\end{aligned}
\eeq
The function $H_L$ is related to the quantity called the generalized Bremsstrahlung function $B_L (\theta)$: 
\beq
H_L(\theta)=\frac{2\theta}{1-\frac{\theta^2}{\pi^2}}B_L(\theta)\period
\eeq
For $L=0$, $B_L (\theta)$ is related to the energy emitted by a moving quark \cite{CHMS} and this is why it is called the generalized Bremsstrahlung function. 
\subsection{Relation to the two-point function}
To compute $B_L$ from our results, one has to relate it to the topological correlators.  For $L=0$ this has already been explained in \cite{CHMS}. As we see below, essentially the same argument applies also to $L\neq 0$  (see also \cite{BGPS}). 

The first step is to consider a small deformation away from the BPS cusp by changing the value of $\theta$. Then, the change of the expectation value can be written as
\beq\label{eq:defWLtheta}
\frac{\delta \langle \mathcal{W}_L\rangle}{\langle \mathcal{W}_L\rangle}=\int_{0}^{\infty} d\tau \,\langle \!\langle \Phi^{\prime}(\tau) \rangle\!\rangle_{\rm cusp}\times \delta\theta 
\eeq
where $\Phi^{\prime}$ is
\beq
\Phi^{\prime}=-\sin \theta \, \Phi_1 + \cos \theta \, \Phi_2\comma
\eeq
and $\langle \!\langle \ast \rangle\!\rangle_{\rm cusp}$ is the normalized correlator of the scalar insertion on the cusped BPS Wilson loop $\mathcal{W}_L (\theta, \phi=\theta)$.
Using the invariance of $\mathcal{W}_L$ under the dilatation around the origin, the $\tau$-dependence of $\langle \!\langle \Phi^{\prime} \rangle\!\rangle_{\rm cusp}$ can be fixed to be 
\beq
\langle \!\langle \Phi^{\prime}(\tau) \rangle\!\rangle_{\rm cusp}=\frac{1}{\tau}\langle \!\langle \Phi^{\prime}(\tau=1) \rangle\!\rangle_{\rm cusp}\period
\eeq
We can then compare \eqref{eq:cuspanodef} with \eqref{eq:defWLtheta} (introducing the UV and IR cutoffs to evaluate the $\tau$ integral), to get
\beq\label{eq:gammaintermsofphi}
\Gamma_{L}= -(\theta-\phi)\langle \!\langle \Phi^{\prime} (\tau=1)\rangle\!\rangle_{\rm cusp}+O((\theta-\phi)^2)\period
\eeq

\begin{figure}
\centering
\includegraphics[clip,height=8cm]{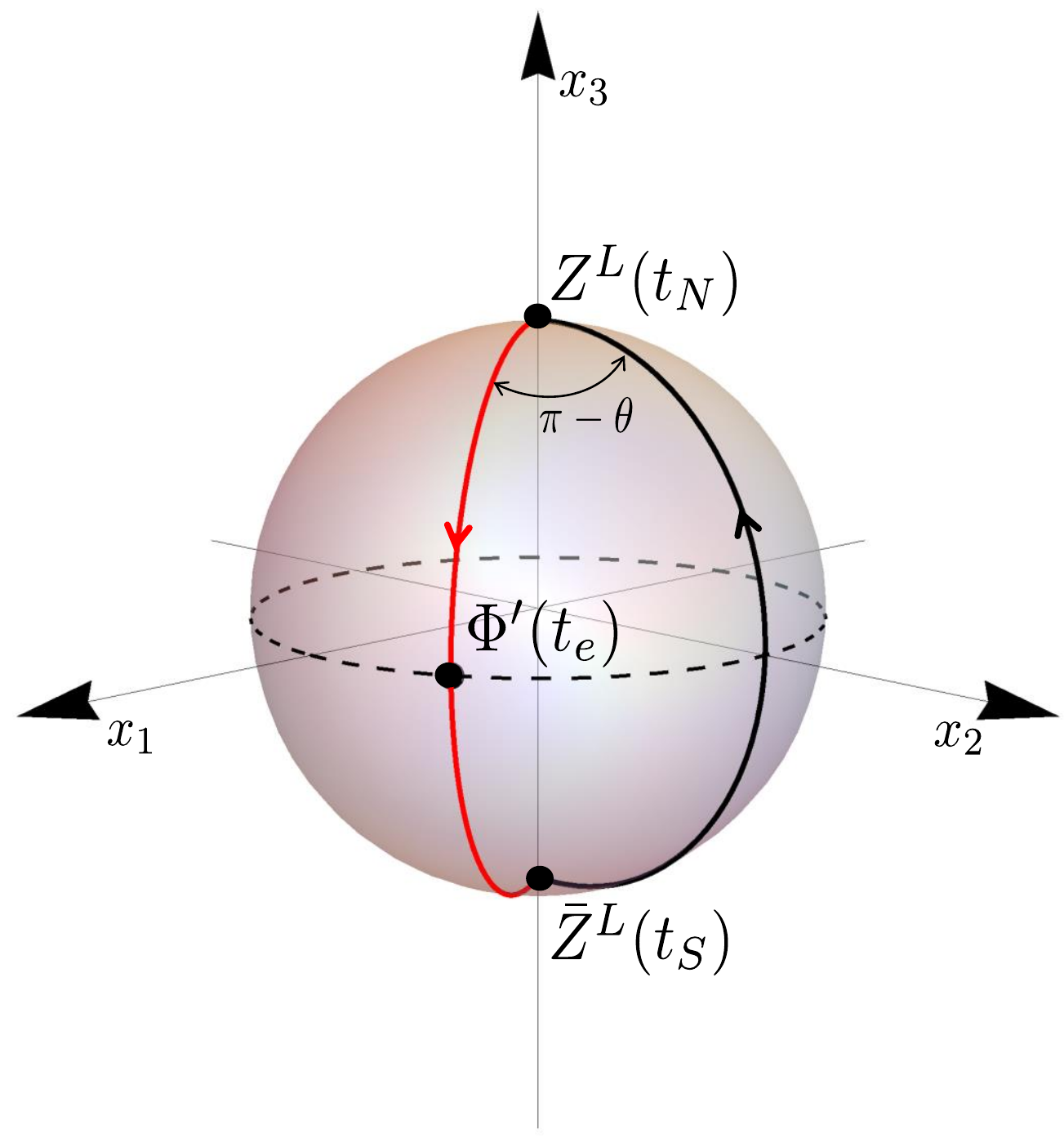}
\caption{Cusped Wilson loop on $S^2$. Applying the conformal transformation, one can map the cusped Wilson line to a configuration depicted above. The red and black semi-circles correspond to the two semi-infinite lines in figure \ref{fig:fig1} of the same color. The angle between the two semi-circles is $\pi-\theta$.  The loop divides the $S^2$ into two regions with areas $2\pi \mp  2\theta$. (Note that we already set $\phi=\theta$ in this figure.) }
\label{fig:cuspS2}
\end{figure}
The second step is to map the BPS cusp $\mathcal{W}_L(\theta,\phi=\theta)$ to the $1/8$ BPS Wilson loop on $S^2$ by the conformal transformation,
\beq
\begin{aligned}
x_1=\frac{2 X_2}{1+X_1^{2}+X_2^{2}} \comma\qquad x_2=\frac{-2 X_1}{1+X_1^{2}+X_2^{2}}\comma\qquad x_3= \frac{1-X_1^{2}-X_2^{2}}{1+X_1^{2}+X_2^{2}}\period
\end{aligned}
\eeq
Here $x_i$'s are the (embedding) coordinates of $S^2$ while $X_i$'s are the coordinates on $R^2$ where the cusped Wilson loop (\ref{cusped-R2}) lives. 
After the transformation, and changing variables by $\tau=-\cot(t/2)$, 
the two semi-infinite lines of the cusped Wilson loop are mapped to the two arcs on $S^2$ (see also figure \ref{fig:cuspS2}),
\beq\label{eq:parametrizationarc}
(x_1,x_2,x_3) =\begin{cases}(0,\sin t, -\cos t)\qquad &0<t\leq \pi\\(-\sin \theta \sin t, \cos \theta \sin t, -\cos t)\qquad &\pi<t\leq 2\pi\comma\end{cases}
\eeq
where the first arc $(0<t\leq \pi)$ and the second arc $(\pi<t\leq 2\pi)$ correspond to the black and the red lines in figure \ref{fig:fig1} 
respectively.  The first arc couples to $\Phi^1$, and the second one to $\cos\theta \Phi^1+\sin\theta \Phi^2$, in accordance 
with our conventions (\ref{eq:defofWL}) for the 1/8-BPS loop. As shown in figure \ref{fig:cuspS2}, the resulting Wilson loop has cusps at the north and the south poles ($t_N$ and $t_S$) with insertions $Z^{L}$ and $\bar{Z}^{L}$ respectively. The insertion $\Phi^{\prime}(\tau=1)$ is mapped to the insertion at a point\fn{In terms of the parametrization given in \eqref{eq:parametrizationarc}, $t_S=0$, $t_N=\pi$ and $t_e=3\pi/2$.} $t_e$ where the red arc intersects the equator of $S^2$. 
We then arrive at the relation between the expectation values,
\beq\label{eq:cuspphi2S2}
\langle \!\langle \Phi^{\prime}(\tau=1) \rangle\!\rangle_{\rm cusp}=\frac{\langle Z^{L}(t_N)\,\,\bar{Z}^{L}(t_S) \,\,\Phi^{\prime}(t_e)\rangle}{\langle Z^{L}(t_N)\,\,\bar{Z}^{L}(t_S)\rangle}
\eeq
where $\langle \ast\rangle$ denotes a (un-normalized) correlator on the Wilson loop on $S^2$. Now, a crucial observation is that one can complete the insertion $\Phi^{\prime} (t_e)$ to the position dependent scalar
\beq
\Phi^{\prime}-i\Phi_4 =-\sin \theta\Phi_1+\cos\theta\Phi_2-i\Phi_4 =-\tilde{\Phi}(t_e)\comma
\eeq
since the correlator with $\Phi_4$ vanishes owing to the charge conservation. Furthermore, $Z^{L}(t_N)$ and $\bar{Z}^{L}(t_S)$ can be identified with $\NO{\tilde{\Phi}^{L}}$. We thus arrive at the following relation\fn{Precisely speaking, the area derivative can also act on the operator $\NO{\tilde{\Phi}^{L}}$ (in addition to inserting an extra single-letter insertion) since it is given by a sum of single-letter insertions with the area-dependent coefficients:
\beq
\NO{\tilde{\Phi}^{L}} =\tilde{\Phi}^{L}+c_1 (A)\tilde{\Phi}^{L-1}+\cdots\period
\eeq
However, since the leading coefficient is $1$, $\del_A \!\!\NO{\tilde{\Phi}^{L}}$ only starts with $\tilde{\Phi}^{L-1}$. Therefore, one can always express $\del_A\!\! \NO{\tilde{\Phi}^{L}}$ as a sum of $\NO{\tilde{\Phi}^{k}}$ with $k<L$. We thus conclude that such contributions vanish because of the orthogonality, $\langle  \NO{\tilde{\Phi}^{k}}\,\, \NO{\tilde{\Phi}^{L}}\rangle=0$ for $k< L$, and do not affect \eqref{eq:relatingcuspwithlogder}. 
 },
\beq\label{eq:relatingcuspwithlogder}
\langle \!\langle \Phi^{\prime} \rangle\!\rangle_{\rm cusp}=-\frac{\langle \NO{\tilde{\Phi}^{L}}\,\,\NO{\tilde{\Phi}^{L}}\tilde{\Phi}\rangle}{\langle \NO{\tilde{\Phi}^{L}}\,\,\NO{\tilde{\Phi}^{L}}\rangle}=-\del_A \log \langle \NO{\tilde{\Phi}^{L}}\,\,\NO{\tilde{\Phi}^{L}}\rangle \period
\eeq
Note that $\tilde{\Phi}$ in the middle is not normal-ordered since it comes directly from the deformation of the loop.
Since the area surrounded by this loop is given by $A=2\pi -2\theta$, one can express $-\del_A$ also as $\del_{\theta}/2$.

From \eqref{eq:gammaintermsofphi} and \eqref{eq:relatingcuspwithlogder}, we can compute the generalized Bremsstrahlung function as
\beq\label{eq:finalgenBrem}
\begin{aligned}
H_L(\theta)=\frac{2\theta}{1-\frac{\theta^2}{\pi^2}}B_L(\theta)=-\frac{1}{2}\del_{\theta} \log \frac{D_{L+1}}{D_L}\comma
\end{aligned}
\eeq
As given in \eqref{eq:halfphiD}, $D_L$ is the following simple determinant, 
\beq
\label{DL-explicit}
D_L = \det{}_{ i,j}\left[(\del_A)^{i+j-2}\langle \mathcal{W}\rangle\right]\qquad (1\leq i,j\leq L)\comma
\eeq
with $A=2\pi -2\theta$. In the limit $\theta \to 0$, the formula takes a particularly simple form,
\beq
B_L(0)=\left.-\frac{1}{4}\del_{\theta}^2\log \frac{D_{L+1}}{D_L}\right|_{\theta=0}=\left.-\del_{A}^2\log \frac{D_{L+1}}{D_L}\right|_{A=2\pi}\period
\label{BLzero}
\eeq 

This is the main result of this section. In the next section, we will see that the formula \eqref{eq:finalgenBrem} reproduces the results in \cite{GS,GLS} in the large $N$ limit. Note that, for $L=1$, our result \eqref{eq:finalgenBrem} is valid also at finite $N$ since we can neglect the mixing with the multi-trace operators (see section \ref{subsec:constructionofoperatorfrom}). The explicit results for $\theta=0$ are given in the Appendix.

\section{Large $N$ limit\label{sec:largeN}}
In this section, we study in detail the topological correlators in the large $N$ limit. In particular, we derive a simple integral expression.
\subsection{Integral expression for topological correlators}
As mentioned before, an important simplification in the large $N$ limit is that $\left< \mathcal{W}\right>$ can be expressed in terms of the deformed Bessel function \eqref{eq:WClargeN}. A nice feature of the deformed Bessel function is that it admits an integral expression \cite{SV},
\beq\label{eq:deformedBessel}
I_n^{\theta}=\oint \frac{dx}{2\pi i x^{n+1}} \sinh(2\pi g (x+1/x))e^{2g \theta (x-1/x)}\,,
\eeq
where here and below we use the notation 
\begin{equation}
g\equiv \frac{\sqrt{\lambda}}{4\pi}\,.
\end{equation}
Applying this to \eqref{eq:WClargeN}, we can express $\left< \mathcal{W}\right>$ and its derivatives simply as
\begin{align}\label{eq:integralW}
&\left< \mathcal{W}\right>=\oint d\mu \comma
\\& \langle \mathcal{W}\rangle ^{(n)}\left(\equiv (\del_A)^{n}\langle \mathcal{W}\rangle\right)=\oint d\mu\,\, \left(g(x-x^{-1})\right)^{n}\comma\label{eq:integralWder}
\end{align}
where the measure $d\mu$ is defined by
\beq\label{eq:defofmumeasure}
d\mu =\frac{dx}{2\pi i x^2} \frac{\sinh(2\pi g (x+1/x))e^{ga (x-1/x)}}{2\pi g}\period
\eeq
Recall that $a=A-2\pi$ and the $1/2$-BPS Wilson loop corresponds to $a=0$. Combining this integral expression with the formula \eqref{eq:generalhpt}, we obtain a simple integral expression for the multi-point correlators,
\beq
\begin{aligned}\label{eq:integraltopologicalcor}
\langle  \NO{\tilde{\Phi}^{L_1}}\NO{\tilde{\Phi}^{L_2}}\cdots \NO{\tilde{\Phi}^{L_m}}\rangle=
\oint d\mu\, \prod_{k=1}^{m}Q_{L_k}(x)\comma
\end{aligned}
\eeq
with 
\beq
Q_{L}(x)\equiv F_{L}\left(g(x-x^{-1})\right)\period
\eeq
At this point, $Q_L$ is just a rewriting of the polynomial $F_L$. However, as we will show in the rest of this section, it is related to the Quantum Spectral Curve \cite{QSC}.
\subsection{Properties of $Q_L(x)$}
The functions $Q_L(x)$ have several important properties.
First, owing to the orthogonality of the two-point functions, they satisfy the following orthogonality relation:
\beq\label{eq:neworth}
\oint d\mu (x)\, Q_L (x)Q_M (x)=\frac{D_{L+1}}{D_{L}}\delta_{LM}\period
\eeq
Second, they are normalized as
\beq\label{eq:QLInormalization}
Q_L(x) =g^{L}x^{L}+\cdots+(-g)^{L}x^{-L}\period
\eeq
Third, since $Q_L(x)$ is a polynomial of $X=g(x-x^{-1})$, it follows that
\beq\label{eq:Qisinvariant}
Q_L(x)=Q_L(-1/x)\period
\eeq
Furthermore, they satisfy the following equalities:
\begin{align}
&\oint d\mu (x)\,x Q_L (x)Q_M (x)=0\comma\label{eq:neworthQ}\\
&\oint d\mu (x)\,x^2 Q_L (x)Q_M (x)=\frac{D_{L+1}}{D_{L}}\delta_{LM}\label{eq:neworthQ2}\period
\end{align}
The first equality follows from 
\beq
\begin{aligned}
\oint d\mu (x)\,x Q_L (x)Q_M (x)&=\oint d\mu (-1/x)\,\left(-\frac{1}{x}\right) Q_L (-1/x)Q_M (-1/x)\\
&=-\oint d\mu (x)\,x Q_L (x)Q_M (x)\comma
\end{aligned}
\eeq
where in the second equality we used the property of the measure\fn{Note that an extra minus sign comes from a change of the direction of the contour.}, $\int d \mu(-1/x)=\int x^2 d\mu (x)$. In a similar manner, the second equality can be proven:
\beq
\begin{aligned}
\oint d\mu (x)\,x^2 Q_L (x)Q_M (x)&=\oint d\mu (-1/x)\,\left(-\frac{1}{x}\right)^2 Q_L (-1/x)Q_M (-1/x)\\
&=\oint d\mu (x)\, Q_L (x)Q_M (x)\period
\end{aligned}
\eeq

The equalities \eqref{eq:neworth}, \eqref{eq:neworthQ} and \eqref{eq:neworthQ2} imply that $Q_L (x)$ and $x Q_L (x)$ for $L\in \mathbb{N}_{\geq 0}$ together form a set of orthogonal functions under the measure $d\mu (x)$. They are in fact the Gram-Schmidt basis obtained by applying the orthogonalization to the set of functions $\{1,x,x^{-1},x^2,x^{-2},\ldots\}$. As we see below, this characterization of the functions $Q_L (x)$ plays a key role in identifying them with the functions introduced in the integrability-based approaches \cite{SV,GLS}.
\subsection{Comparison with the results from integrability\label{subsec:comparison}}
We now prove the equivalence between our results and the results obtained previously from integrability \cite{GS,GLS}. For this purpose, we first show that $Q_L$ coincides with the function $P_L$, which was introduced in \cite{GS,GLS} and later shown to be directly related to the so-called ``$Q$-functions'' in the Quantum Spectral Curve\cite{GL}. The equivalence of other quantities, including the generalized Bremsstrahlung functions, follow from it.

As the first step, let us recall the polynomials $P_L$ defined in \cite{GLS}:
\beq
\begin{aligned}
P_L(x)\equiv \frac{1}{m_{2L}}\dmatrix{ccccc}{I_1^{\theta}&I_0^{\theta}&\cdots &I_{2-2L}^{\theta}&I_{1-2L}^{\theta}\\
I_2^{\theta}&I_1^{\theta}&\cdots &I_{3-2L}^{\theta}&I_{2-2L}^{\theta}\\
\vdots&\vdots&\ddots &\vdots&\vdots\\
I_{2L}^{\theta}&I_{2L-1}^{\theta}&\cdots &I_{1}^{\theta}&I_{0}^{\theta}\\
x^{-L}&x^{1-L}&\cdots&x^{L-1}&x^L}\comma\qquad L\geq 0\comma
\end{aligned}
\eeq
where one has to set $\theta =a/2$ to compare with our formulae and $m_{L}$ is given by
\beq
\begin{aligned}
m_{L}=\dmatrix{cccc}{I_1^{\theta}&I_0^{\theta}&\cdots &I_{2-L}^{\theta}\\
I_2^{\theta}&I_1^{\theta}&\cdots &I_{3-L}^{\theta}\\
\vdots&\vdots&\ddots &\vdots\\
I_{L}^{\theta}&I_{L-1}^{\theta}&\cdots &I_{1}^{\theta}}\period
\end{aligned}
\eeq
Note that we changed the normalization of $P_L$ slightly so that the coefficient of the leading term becomes unity\fn{
The relation between the conventions here and the conventions in \cite{GLS,SV,GL} can be summarized as follows:
\beq
\begin{aligned}
&\det \mathcal{M}_{2L}=m_{2L+1}\comma\\
&P_L(x){}_\text{ in \cite{GLS,SV,GL}} =\frac{m_{2L}}{m_{2L+1}} P_L (x){}_{\text{ here}}\left(=\frac{\det \mathcal{M}_{2L-1}}{\det \mathcal{M}_{2L}}P_L(x){}_{\text{ here}}\right)\period
\end{aligned}
\eeq
},
\beq\label{eq:PLnormalization}
P_L(x)=x^{L}+\cdots+\frac{\tilde{m}_{2L}}{m_{2L}}x^{-L}\comma
\eeq
with
\beq
\tilde{m}_{n}=\dmatrix{cccc}{I_0^{\theta}&I_{-1}^{\theta}&\cdots &I_{1-n}^{\theta}\\
I_1^{\theta}&I_0^{\theta}&\cdots &I_{2-n}^{\theta}\\
\vdots&\vdots&\ddots &\vdots\\
I_{n-1}^{\theta}&I_{n-2}^{\theta}&\cdots &I_{0}^{\theta}}\period
\eeq
 
 Let us also introduce another set of functions $\tilde{P}_L(x)$ defined by
 \beq
 \tilde{P}_L(x)\equiv \frac{1}{m_{2L+1}}\dmatrix{ccccc}{I_1^{\theta}&I_0^{\theta}&\cdots &I_{1-2L}^{\theta}&I_{-2L}^{\theta}\\
I_2^{\theta}&I_1^{\theta}&\cdots &I_{2-2L}^{\theta}&I_{1-2L}^{\theta}\\
\vdots&\vdots&\ddots &\vdots&\vdots\\
I_{2L+1}^{\theta}&I_{2L}^{\theta}&\cdots &I_{1}^{\theta}&I_{0}^{\theta}\\
x^{-L}&x^{1-L}&\cdots&x^{L}&x^{L+1}}\comma\qquad L\geq 0\period
 \eeq
 These functions satisfy the following important orthogonality properties:
\begin{align}
&\oint d\mu (x) P_L (x) P_M (x) =\frac{1}{2\pi g}\frac{\tilde{m}_{2L}m_{2L+1}}{m_{2L}^2}\delta_{LM}\comma\label{eq:orthPL}
\\
&\oint d\mu (x)  \tilde{P}_L (x) P_M (x) =0\comma\label{eq:orthPL2}\\
&\oint d\mu (x) \tilde{P}_L (x) \tilde{P}_M (x) =-\frac{1}{2\pi g}\frac{\tilde{m}_{2L+2}}{m_{2L+1}} \delta_{LM}\period\label{eq:orthPL3}
\end{align}
 In what follows, we will prove these relations one by one. 
 
 Let us first consider \eqref{eq:orthPL}. To prove it, it is enough to study the case with $L\geq M$. We first perform the Laplace expansion for $P_M (x)$ to get
\beq
P_M (x)= \sum_{k=- M }^{ M }c_k x^{k}\comma
\eeq
 where $c_k$'s are constants with $c_{M}=1$ and $c_{-M}=\tilde{m}_{2M}/m_{2M}$. Substituting this expression to the left hand side of \eqref{eq:orthPL}, one gets
 \beq
 \begin{aligned}
 \oint d\mu (x) P_L (x) P_M (x) =\sum_{k=- M }^{ M } \frac{c_k}{m_{2L}} \oint d\mu (x) \dmatrix{ccccc}{I_1^{\theta}&I_0^{\theta}&\cdots &I_{2-2L}^{\theta}&I_{1-2L}^{\theta}\\
I_2^{\theta}&I_1^{\theta}&\cdots &I_{3-2L}^{\theta}&I_{2-2L}^{\theta}\\
\vdots&\vdots&\ddots &\vdots&\vdots\\
I_{2L}^{\theta}&I_{2L-1}^{\theta}&\cdots &I_{1}^{\theta}&I_{0}^{\theta}\\
x^{- L +k}&\cdots &\cdots&\cdots &x^{ L +k}}
 \end{aligned}
 \eeq
 Using the integral expression for the deformed Bessel function \eqref{eq:deformedBessel}, which can be expressed in terms of $d\mu$ as
 \beq\label{eq:besselintegralmu}
 I_{n}^{\theta}=2\pi g\oint \frac{d\mu}{x^{n-1}}\comma
 \eeq
 one can perform the integral to get
 \beq
 \begin{aligned}
 \oint d\mu (x) P_L (x) P_M (x) =\frac{1}{2\pi g}\sum_{k=- M }^{ M } \frac{c_k}{m_{2L}}  \dmatrix{ccccc}{I_1^{\theta}&I_0^{\theta}&\cdots &I_{2-2L}^{\theta}&I_{1-2L}^{\theta}\\
I_2^{\theta}&I_1^{\theta}&\cdots &I_{3-2L}^{\theta}&I_{2-2L}^{\theta}\\
\vdots&\vdots&\ddots &\vdots&\vdots\\
I_{2L}^{\theta}&I_{2L-1}^{\theta}&\cdots &I_{1}^{\theta}&I_{0}^{\theta}\\
I_{ L  -k +1}^{\theta}&\cdots &\cdots&\cdots &I_{- L  -k +1}^{\theta}}\period
 \end{aligned}
 \eeq
 If $L>M$, the last row always coincides with one of the rows above. Therefore, all the terms in the sum vanishes and one has
 \beq
 \oint d\mu (x) P_L (x) P_M (x) = 0\comma  \qquad L>M\period 
 \eeq
 On the other hand, for $L=M$, there is one term in the sum which is nonzero: $k=-L$. We thus have
 \beq
 \oint d\mu (x) P_L (x) P_L (x) =\frac{1}{2\pi g}\frac{c_{-L}m_{2L+1}}{m_{2L}}=\frac{1}{2\pi g}\frac{\tilde{m}_{2L}m_{2L+1}}{m_{2L}^2}\period
 \eeq
 
Let us next consider \eqref{eq:orthPL2}. Expanding again the determinant expression for $P_M$ and substituting it to the left hand side of \eqref{eq:orthPL2}, we obtain
  \beq
 \begin{aligned}
 \oint d\mu (x) \tilde{P}_L (x) P_M (x) &=\sum_{k=- M }^{ M } \frac{c_k}{m_{2L+1}} \oint d\mu (x) \dmatrix{ccccc}{I_1^{\theta}&I_0^{\theta}&\cdots &I_{1-2L}^{\theta}&I_{-2L}^{\theta}\\
I_2^{\theta}&I_1^{\theta}&\cdots &I_{2-2L}^{\theta}&I_{1-2L}^{\theta}\\
\vdots&\vdots&\ddots &\vdots&\vdots\\
I_{2L+1}^{\theta}&I_{2L}^{\theta}&\cdots &I_{1}^{\theta}&I_{0}^{\theta}\\
x^{- L +k}&\cdots &\cdots&\cdots &x^{ L +k+1}}\\
&=\frac{1}{2\pi g}\sum_{k=- M }^{ M } \frac{c_k}{m_{2L+1}}  \dmatrix{ccccc}{I_1^{\theta}&I_0^{\theta}&\cdots &I_{1-2L}^{\theta}&I_{-2L}^{\theta}\\
I_2^{\theta}&I_1^{\theta}&\cdots &I_{2-2L}^{\theta}&I_{1-2L}^{\theta}\\
\vdots&\vdots&\ddots &\vdots&\vdots\\
I_{2L+1}^{\theta}&I_{2L}^{\theta}&\cdots &I_{1}^{\theta}&I_{0}^{\theta}\\
I_{ L  -k+1}^{\theta}&\cdots &\cdots&\cdots &I_{- L  -k }^{\theta}}\period
 \end{aligned}
 \eeq
 One can easily see that, for $L\geq M$, the last row coincides with one of the rows above and therefore the sum always vanishes. This proves \eqref{eq:orthPL2} for $L\geq M$. One can also show \eqref{eq:orthPL2} for $L<M$ by performing the Laplace expansion of $\tilde{P}_L$:
 \beq
 \tilde{P}_L (x)=\sum_{k=-L}^{L+1}\tilde{c}_k x^{k}\period
 \eeq
 Substituting this expression to the left hand side of \eqref{eq:orthPL2}, we get
 \beq
 \begin{aligned}
 \oint d\mu (x) \tilde{P}_L (x) P_M (x) &=\sum_{k=- L }^{ L+1 } \frac{\tilde{c}_k}{m_{2M}} \oint d\mu (x) \dmatrix{ccccc}{I_1^{\theta}&I_0^{\theta}&\cdots &I_{2-2M}^{\theta}&I_{1-2M}^{\theta}\\
I_2^{\theta}&I_1^{\theta}&\cdots &I_{3-2M}^{\theta}&I_{2-2M}^{\theta}\\
\vdots&\vdots&\ddots &\vdots&\vdots\\
I_{2M}^{\theta}&I_{2M-1}^{\theta}&\cdots &I_{1}^{\theta}&I_{0}^{\theta}\\
x^{- M +k}&\cdots &\cdots&\cdots &x^{ M +k}}\\
&=\frac{1}{2\pi g}\sum_{k=- L }^{ L+1 } \frac{\tilde{c}_k}{m_{2M}}  \dmatrix{ccccc}{I_1^{\theta}&I_0^{\theta}&\cdots &I_{2-2M}^{\theta}&I_{1-2M}^{\theta}\\
I_2^{\theta}&I_1^{\theta}&\cdots &I_{3-2M}^{\theta}&I_{2-2M}^{\theta}\\
\vdots&\vdots&\ddots &\vdots&\vdots\\
I_{2M}^{\theta}&I_{2N-1}^{\theta}&\cdots &I_{1}^{\theta}&I_{0}^{\theta}\\
I_{ M  -k +1}^{\theta}&\cdots &\cdots&\cdots &I_{- M  -k +1}^{\theta}}\period
 \end{aligned}
 \eeq
 Again, one can show that the last row always coincides with one of the rows above as long as $L<M$. This completes the proof of \eqref{eq:orthPL2}.
 
 Let us finally show \eqref{eq:orthPL3}. Again it is enough to consider the case with $L\geq M$. By performing the Laplace expansion of $\tilde{P}_M$ and substituting it to \eqref{eq:orthPL3}, one gets
 \beq
\begin{aligned}
 \oint d\mu (x) \tilde{P}_L (x) \tilde{P}_M (x) &=\sum_{k=- M }^{ M+1 } \frac{\tilde{c}_k}{m_{2L+1}} \oint d\mu (x) \dmatrix{ccccc}{I_1^{\theta}&I_0^{\theta}&\cdots &I_{1-2L}^{\theta}&I_{-2L}^{\theta}\\
I_2^{\theta}&I_1^{\theta}&\cdots &I_{2-2L}^{\theta}&I_{1-2L}^{\theta}\\
\vdots&\vdots&\ddots &\vdots&\vdots\\
I_{2L+1}^{\theta}&I_{2L}^{\theta}&\cdots &I_{1}^{\theta}&I_{0}^{\theta}\\
x^{- L +k}&\cdots &\cdots&\cdots &x^{ L +k+1}}\\
&=\frac{1}{2\pi g}\sum_{k=- M }^{ M+1 } \frac{\tilde{c}_k}{m_{2L+1}}  \dmatrix{ccccc}{I_1^{\theta}&I_0^{\theta}&\cdots &I_{1-2L}^{\theta}&I_{-2L}^{\theta}\\
I_2^{\theta}&I_1^{\theta}&\cdots &I_{2-2L}^{\theta}&I_{1-2L}^{\theta}\\
\vdots&\vdots&\ddots &\vdots&\vdots\\
I_{2L+1}^{\theta}&I_{2L}^{\theta}&\cdots &I_{1}^{\theta}&I_{0}^{\theta}\\
I_{ L  -k+1}^{\theta}&\cdots &\cdots&\cdots &I_{- L  -k }^{\theta}}\period
 \end{aligned} 
 \eeq
 For $L>M$, the determinant always vanishes for the same reason as the previous discussions. On the other hand, if $L=M$, the term with $k=L+1$ does not vanish and gives\fn{The minus sign comes from the reordering of the matrix.}
 \beq
 \oint d\mu (x) \tilde{P}_L (x) \tilde{P}_L (x) =-\frac{\tilde{c}_{L+1}\tilde{m}_{2L+2}}{(2\pi g)m_{2L+1}}=-\frac{1}{2\pi g}\frac{\tilde{m}_{2L+2}}{m_{2L+1}}\period
 \eeq
 
 Now, from the relations \eqref{eq:orthPL}, \eqref{eq:orthPL2} and \eqref{eq:orthPL3} together with the normalization \eqref{eq:PLnormalization}, it follows that the set of functions $\{ P_L (x), \tilde{P}_L(x)\}$ forms a Gram-Schmidt basis obtained from the functions $\{1,x,x^{-1},x^{2},x^{-2},\ldots\}$. Since the Gram-Schmidt basis is unique up to overall normalizations, we conclude that $\{ P_L (x), \tilde{P}_L(x)\}$ must be proportional to $\{Q_L(x), xQ_L(x)\}$. The constants of proportionality can be fixed by comparing \eqref{eq:QLInormalization} and \eqref{eq:PLnormalization} and we arrive at
 \beq\label{eq:finalQP}
 P_L (x)=\frac{Q_L(x)}{g^{L}}\comma\qquad \tilde{P}_L(x)=\frac{xQ_L(x)}{g^{L}}\period
 \eeq 
 This in particular means that $P_L(x)$ also has a property
 \beq
 P_L (x)=P_L(-1/x)\period
 \eeq
 Imposing this property on the expansion \eqref{eq:PLnormalization}, we get
 \beq
 (-1)^{L}=\frac{\tilde{m}_{2L}}{m_{2L}}\period
 \eeq
 We can thus rewrite the relations \eqref{eq:orthPL} and \eqref{eq:orthPL3} as
\begin{align}
&\oint d\mu (x) P_L (x) P_M (x) =(-1)^L\frac{1}{2\pi g}\frac{m_{2L+1}}{m_{2L}}\delta_{LM}\comma\label{eq:neworthPL}
\\
&\oint d\mu (x) \tilde{P}_L (x) \tilde{P}_M (x) =(-1)^{L}\frac{1}{2\pi g}\frac{m_{2L+2}}{m_{2L+1}} \delta_{LM}\period\label{eq:neworthPL2}
\end{align}
Comparing these relations with \eqref{eq:neworth} and \eqref{eq:neworthQ2} in view of the correspondence \eqref{eq:finalQP}, we obtain the relation between the ratios of determinants
\beq
\frac{D_{L+1}}{D_L}= (-1)^{L}\frac{g^{2L-1}}{2\pi}\frac{m_{2L+1}}{m_{2L}}=(-1)^{L}\frac{g^{2L-1}}{2\pi}\frac{m_{2L+2}}{m_{2L+1}}\comma
\eeq
which leads to
\beq
\frac{g^{4L-2}}{(2\pi)^2}\frac{m_{2L+2}}{m_{2L}}=\left(\frac{D_{L+1}}{D_L}\right)^2\period
\eeq
Using the initial condition $m_0=D_0=1$, we can solve the recursion to get
\beq\label{eq:relationmMdM}
m_{2L}=\frac{(2\pi)^{2L}}{g^{2L(L-2)}} (D_L)^2\period
\eeq
This establishes the relation between the  two determinant expressions.

Using this relation, we can express the large $N$ limit of the generalized Bremsstrahlung function \eqref{eq:finalgenBrem} as
\beq\label{eq:finalgenBrem2}
\begin{aligned}
H_L(\theta)=\frac{2\theta}{1-\frac{\theta^2}{\pi^2}}B_L(\theta)=-\frac{1}{4}\del_{\theta} \log \frac{m_{2L+2}}{m_{2L}}\period
\end{aligned}
\eeq
 This is precisely the result obtained previously from integrability \cite{GLS}.
\subsection{Variations of the measure}
Before proceeding, let us now make a small remark on the measure $d\mu$. The expectation value of the Wilson loop admits several different integral representations besides \eqref{eq:integralW}:
\beq
\begin{aligned}
\langle \mathcal{W}\rangle =\oint d\mu_{\rm sym}=\oint d\mu_{\rm exp}\period
\end{aligned}
\eeq
Here $d\mu_{\rm sym}$ is a ``symmetrized'' measure defined by
\beq\label{eq:symmeasure}
d\mu_{\rm sym}=\frac{dx}{2\pi i}\frac{1+x^{-2}}{2}\frac{\sinh (2\pi g(x+1/x))e^{ga(x-1/x)}}{2\pi g}\comma
\eeq
while $d\mu_{\rm exp}$ is an ``exponential'' measure defined by
\beq\label{eq:expmeasure}
d\mu_{\rm exp}=\frac{dx}{2\pi i}\frac{1+x^{-2}}{2}\frac{e^{2\pi g(x+1/x)}e^{ga(x-1/x)}}{2\pi g}\period
\eeq
The symmetrized expression can be derived from the original one \eqref{eq:integralW} by performing the transformation $x\to -1/x$ and averaging the original expression and the transformed one. On the other hand, the exponential expression can be obtained from the symmetrized one by splitting $\sinh$ into two exponentials and performing $x\to -1/x$ to $e^{-2\pi g (x+1/x)}$.

Almost all the results obtained so far in this section are valid even if we replace $d\mu$ with $d\mu_{\rm sym}$ or $d\mu_{\rm exp}$ since the functions $Q_L(x)$ are invariant under $x\to -1/x$; see \eqref{eq:Qisinvariant}. (The only exceptions are \eqref{eq:neworthQ} and \eqref{eq:neworthQ2} whose derivation relies crucially on the property of $d\mu$.) This in particular means that one can alternatively use $d\mu_{\rm sym}$ or $d\mu_{\rm exp}$ for the integral expression for the topological correlators \eqref{eq:integraltopologicalcor}. In the following sections, we will see the uses of these other measures.  
\subsection{Nonplanar corrections to the measure}\label{subsec:nonplanar}
So far, we have been discussing the large $N$ limit in this section. As we explain below, it is also possible to incorporate some of the non-planar effects, namely the non-planar corrections to $\langle \mathcal{W}\rangle$, into the measure factor.

The large $N$ expansion of the expectation value of the Wilson loop is given by \cite{DG}
\beq
\langle \mathcal{W}\rangle=\frac{2}{\sqrt{\lambda^{\prime}}}I_1 (\sqrt{\lambda^{\prime}})+\frac{\lambda^{\prime}}{48N^2}I_2 (\sqrt{\lambda^{\prime}})+\cdots\comma
\eeq
with $\lambda^{\prime}\equiv \lambda \left(1-a^2/(4\pi^2)\right)=(4\pi g)^2\left(1-a^2/(4\pi^2)\right)$.
To find the first non-planar correction to the measure, we use the generating function\fn{This follows from the usual generating function for the modified Bessel function, $e^{\frac{\sqrt{\lambda^{\prime}}}{2}(y+\frac{1}{y})}=\sum_{n}I_n (\sqrt{\lambda^{\prime}}) y^{n}$, after the change of variables $y=\sqrt{\frac{2\pi+a}{2\pi-a}}\,\,x$.},
\beq
e^{2\pi g(x+\frac{1}{x})}e^{ga(x-\frac{1}{x})}=\sum_{n=-\infty}^{+\infty}I_{n}(\sqrt{\lambda^{\prime}}) \left(\frac{2\pi +a}{2\pi -a}\right)^{n/2}x^n\period
\eeq
which leads to the following integral representation:
\beq
I_{n}(\sqrt{\lambda^{\prime}}) \left(\frac{2\pi +a}{2\pi -a}\right)^{n/2}=\oint \frac{dx}{2\pi i x^{n+1}}e^{2\pi g(x+\frac{1}{x})}e^{ga(x-\frac{1}{x})}
\eeq
Applying this to the first nonplanar correction, we get
\beq\label{eq:I2integral}
\begin{aligned}
\frac{\lambda^{\prime}}{48N^2}I_2 (\sqrt{\lambda^{\prime}}) &=\frac{\lambda^{\prime}}{48N^2}\frac{2\pi-a}{2\pi+a}\oint \frac{dx}{2\pi i x^{3}}e^{2\pi g(x+\frac{1}{x})}e^{a(x-\frac{1}{x})}\\
&=\frac{g^2(2\pi -a)^2}{12N^2}\oint \frac{dx}{2\pi i x^{3}}e^{2\pi g(x+\frac{1}{x})}e^{a(x-\frac{1}{x})}
\end{aligned}
\eeq
The expression \eqref{eq:I2integral} contains an extra dependence on the area, $(2\pi -a)^2$. However, this can be absorbed into the integral by using
\beq
\begin{aligned}
&ga \times e^{ga(x-\frac{1}{x})}=\frac{1}{1+1/x^2}\frac{d e^{ga(x-\frac{1}{x})}}{dx}\comma\\
&(ga)^2\times e^{ga(x-\frac{1}{x})}=\frac{1}{1+1/x^2}\frac{d}{dx}\left[\frac{1}{1+1/x^2}\frac{d e^{ga(x-\frac{1}{x})}}{dx}\right]
\end{aligned}
\eeq
and performing the integration by parts. As a result, we get
\beq
\begin{aligned}
\frac{\lambda^{\prime}}{48N^2}I_2 (\sqrt{\lambda^{\prime}}) &
&=\oint \frac{dx}{2\pi i x}\,\,e^{2\pi g(x+\frac{1}{x})}e^{ga(x-\frac{1}{x})}\,\, f(2\pi g (x+1/x))\comma
\end{aligned}
\eeq
with
\beq
f(z)=\frac{(2\pi g)^4}{N^2}\frac{z^2-3z+3}{3z^4}\period
\eeq
Thus, using the exponential measure for the planar part, one can write down the corrected measure $d\mu_{1/N}$ as
\beq
d\mu_{1/N}=\frac{dx}{2\pi i x}\,\,e^{2\pi g(x+\frac{1}{x})}e^{ga(x-\frac{1}{x})}\,\, F(2\pi g (x+1/x))\comma
\eeq
with
\beq
F(z)=\frac{1}{(2\pi g)^2}\frac{z}{2}+\frac{(2\pi g)^4}{N^2}\frac{z^2-3z +3}{3z^4}+O(1/N^{4})\cdots\period
\eeq
Since the area dependence only appears in the exponent $e^{ga(x-1/x) }$,  the expectation value of the Wilson loop and its derivatives retain the following simple expressions:
\beq
\langle \mathcal{W}\rangle^{(n)} =\oint d\mu_{1/N} \left(g(x-x^{-1})\right)^{n}\period 
\eeq 
Thanks to this property, one can compute the non-planar corrections discussed in this subsection by simply replacing the measure $d\mu$ to $d\mu_{1/N}$ in the integral expression for the topological correlators \eqref{eq:integraltopologicalcor}. Note however that the correction discussed in this subsection only captures a part of the full non-planar correction since there are additional contributions coming from the mixing with the multi-trace operators as discussed in section \ref{subsec:constructionofoperatorfrom}.

Repeating the same analysis at higher orders, we can determine the corrections to the measure order by order. After working out first several orders, we found\fn{We only checked the relation by Mathematica and did not work out a proof. It would be nice to prove and establish the relation.} the following relation between the terms that appear in the expansion of $\langle \mathcal{W}\rangle$ and the corrections to $F(z)$:
\beq
(\lambda^{\prime})^{\frac{n}{2}}I_{n}(\sqrt{\lambda^{\prime}})\quad \overset{\text{Integration by parts}}{\longrightarrow} \quad \left[\frac{(4\pi g)^2}{z}\right]^{n}\sqrt{\frac{2}{\pi}}e^{-z}\sqrt{-z} K_{n+\frac{1}{2}}(-z)
\eeq
Note that although the right hand side involves the modified Bessel function $K_{n+\frac{1}{2}}$, it actually reduces to a rational function of $z$. Now, applying this relation to the expansion of $\langle \mathcal{W} \rangle$ given by Drukker and Gross \cite{DG},
\beq\label{eq:DGexp}
\langle \mathcal{W}\rangle =\frac{2}{\sqrt{\lambda^{\prime}}}I_{1}(\sqrt{\lambda^{\prime}})+\sum_{k=1}^{\infty}\frac{1}{N^{2k}}\sum_{s=0}^{k-1}X_k^{s}\left(\frac{\lambda^{\prime}}{4}\right)^{\frac{3k-s-1}{2}}I_{3k-s-1}(\sqrt{\lambda^{\prime}})\comma
\eeq
we obtain the following expansion of $F(z)$:
\beq\label{eq:DGexp2}
F(z)=\frac{1}{(2\pi g)^2}\frac{z}{2}+\sqrt{\frac{-2z}{\pi}}e^{-z}\sum_{k=1}^{\infty}\frac{1}{N^{2k}}\sum_{s=0}^{k-1}X_{k}^{s}\left[\frac{(2\pi g)^2}{z}\right]^{(3k-s-1)} K_{(3k-s-1)+\frac{1}{2}}(-z)
\eeq
In \eqref{eq:DGexp} and \eqref{eq:DGexp2}, $X_{k}^{s}$ is a numerical coefficient defined by the following recursion:
\beq
\begin{aligned}
&4X_k^{s}=\frac{3k-s-2}{3k-s}X^{s-1}_{k-1}+\frac{1}{3k-s}X^{s}_{k-1}\comma\\
&X_{1}^{0}=\frac{1}{12}\comma\qquad X_{k}^{k}=0\period
\end{aligned}
\eeq
It would be interesting to try to resum the series \eqref{eq:DGexp2} and to consider the nonperturbative corrections.
\section{Weak- and strong-coupling expansions\label{sec:weakstrong}}
We now discuss the weak- and the strong-coupling expansions of topological correlators on the $1/2$-BPS Wilson loop at large $N$, and compare them with the direct perturbative results. In particular, we focus on the three-point functions since the topological correlators are closed under the OPE and the match of the three-point functions (or equivalently the OPE coefficients) automatically guarantee the match of higher-point functions. 

In both cases, we first compute the expansion of the polynomials $Q_L(x)$:
\beq
Q_L(x)=\begin{cases}Q_L^{0}(x)+g^2Q^{1}_{L}(x)+\cdots\quad &g\ll 1\\\bar{Q}_L^{0}(x)+\frac{1}{g}\bar{Q}^{1}_{L}(x)+\cdots\quad &g\gg 1\end{cases}\period
\eeq
To determine the expansion, it is convenient to use the symmetrized measure $d\mu_{\rm sym}$ \eqref{eq:symmeasure} and perform the change of variables from $x$ to $y\equiv i(x-x^{-1})/2$. Then, the integral over $x$ can be rewritten as the following integral of $y$:
\beq\label{eq:goodtoexp}
\oint d\mu_{\rm sym}\left(\cdots\right)=\frac{2}{\pi}\int_{-1}^{1}dy \,\frac{\sinh (4\pi g \sqrt{1-y^2})}{4\pi g}\left(\cdots\right)\period
\eeq
Note that we set $a=0$ since we consider the $1/2$-BPS loops.
In what follows, we use this representation for the measure to compute the weak- and the strong-coupling expansions.
\subsection{Weak coupling expansion}
Let us expand the measure \eqref{eq:goodtoexp} at weak coupling,
\beq
 d\mu_{\rm sym}=d\mu^{0}+g^2d\mu^{1}+O(g^4)\cdots\period
\eeq
At the leading order, it is given by 
\beq
d\mu^{0}=\frac{2}{\pi} dy \sqrt{1-y^2}\period
\eeq
This coincides with the measure for the Chebyshev polynomials of the second kind. Thus, taking into account the difference of the normalization, we conclude that $Q_{L}$ at the leading order at weak coupling is given by
\beq
Q_L^{0}(x)=(-ig)^{L}U_L (y)\comma
\eeq
where $U_L (y)$ is the Chebyshev polynomial of the second kind determined by the following recursion relation:
\beq
\begin{aligned}
&U_0 (y)=1\comma \qquad U_1 (y)=2y\comma\\
&U_{L+1}(y)=2y U_{L}(y)-U_{L-1}(y)\period
\end{aligned}
\eeq

Having identified $Q_N$ with the Chebyshev polynomial, one can now compute the two- and the three-point functions by using the identities,
\beq
\begin{aligned}
&\frac{\pi }{2}\int dy \sqrt{1-y^2}U_{L}(y)U_M(y)=\delta_{LM}\comma\\
&U_L(y)U_M(y)=\sum_{k=0}^{M}U_{L-M+2k}(y)\qquad (L\geq M)\period
\end{aligned}
\eeq
Using these identities to evaluate the integral expressions for the correlators \eqref{eq:integraltopologicalcor}, we get
\beq
\begin{aligned}
&\left.\langle  \NO{\tilde{\Phi}^{L_1}}\NO{\tilde{\Phi}^{L_2}}\rangle\right|_{O(g^0)}= (-g^2)^{L_1}\delta_{L_1,L_2}\comma\\
&\left.\langle  \NO{\tilde{\Phi}^{L_1}}\NO{\tilde{\Phi}^{L_2}}\NO{\tilde{\Phi}^{L_3}}\rangle\right|_{O(g^0)}=(-g^2)^{\frac{L_{\rm tot}}{2}}d_{L_1,L_2,L_3}\comma
\end{aligned}
\eeq
where $L_{\rm tot}$ is given by
\beq
L_{\rm tot}\equiv L_1+L_2+L_3\comma
\eeq
and the symbol $d_{L_1,L_2,L_3}$ denotes
\beq\label{eq:defofdfunctions}
d_{L_1,L_2,L_3} =\begin{cases}1 \qquad \left(L_i+L_j\geq L_k\right)\wedge \left(\sum_{s=1}^{3}L_s:{\rm even}\right)\\0\qquad {\rm otherwise}\end{cases} \period
\eeq
As shown in \eqref{eq:defofdfunctions} the three-point function is nonzero only when the triangle inequalities are satisfied and the sum of the lengths of the operators is even. These results precisely match the tree-level planar Wick contractions. Note that, at this order, the expectation value of the Wilson loop is $1$ and there is no distinction between the un-normalized and the normalized correlators $\langle\!\langle \ast \rangle\!\rangle$.

Let us now discuss the one-loop correction. At one loop, the measure receives an additional contribution,
\beq\label{eq:dmu1weak}
d\mu^{1}= d\mu^{0}\times \frac{8\pi^2}{3}(1-y^2)
\eeq
This change of the measure induces the change of the orthogonal polynomials $Q_L^{1}$ since they need to satisfy the modified orthogonality condition
\beq\label{eq:modifiedweak}
\int  d\mu^{1} Q^{0}_{L}Q_{M}^{0}+ \int d\mu^{0} Q^{1}_{L}Q_{M}^{0}+\int d\mu^{0} Q^{0}_{L}Q_{M}^{1}\propto \delta_{LM}\period
\eeq
Furthermore, in order to keep the normalization condition \eqref{eq:QLInormalization}, the correction $Q_L^{1}$ must be a polynomial of $y$ with the order $<L$. One can solve these conditions  using the equality
\beq
(1-y^2)U_L(y)=\frac{1}{4}\left(2U_L (y)-U_{L+2}(y)-U_{L-2}(y)\right)\comma
\eeq
and the result reads
\beq
Q_L^{1}(y)=(-ig)^{L}\frac{2\pi^2}{3}U_{L-2}(y)\period
\eeq
We can then compute the correction to the two- and the three-point functions using the integral representation for the correlators \eqref{eq:integraltopologicalcor} as follows:
\beq
\begin{aligned}
\left.\langle  \NO{\tilde{\Phi}^{L_1}}\NO{\tilde{\Phi}^{L_2}}\rangle\right|_{O(g^2)}=&\frac{(2\pi g)^2(-g^2)^{L_1}}{3} \delta_{L_1,L_2}\comma\\
\left.\langle  \NO{\tilde{\Phi}^{L_1}}\NO{\tilde{\Phi}^{L_2}}\NO{\tilde{\Phi}^{L_3}}\rangle\right|_{O(g^2)}=&\frac{(2\pi g)^2(-g^2)^{\frac{L_{\rm tot}}{2}}}{6} (2d_{L_1,L_2,L_3}+d_{L_1,L_2-2,L_3}\\
&\qquad\qquad \qquad \qquad +d_{L_1,L_2,L_3-2}-d_{L_1+2,L_2,L_3})\comma
\end{aligned}
\eeq
where $d_{a,b,c}$ is 1 only when $a+b+c$ is even and they satisfy the triangular inequality, (otherwise zero). Using the identities\fn{These identities can be derived by expressing $d_{a,b,c}$ as a product of step functions
\beq
\Theta (x)=\begin{cases}1\qquad (x\geq 0)\\0\qquad (x<0)\end{cases}
\eeq and using the fact that $\Theta (x+1)= \Theta (x)+\delta_{x,-1}$ and $\Theta (x-1)=\Theta (x)-\delta_{x,0}$.}
\beq
\begin{aligned}
&d_{L_1,L_2-2,L_3}=d_{L_1,L_2,L_3}-\delta_{L_2+L_3,L_1}-\delta_{L_1+L_2,L_3}+\delta_{L_3+L_1,L_2-2}\comma\\
&d_{L_1,L_2,L_3-2}=d_{L_1,L_2,L_3}-\delta_{L_3+L_1,L_2}-\delta_{L_2+L_3,L_1}+\delta_{L_1+L_2,L_3-2}\comma\\
&d_{L_1+2,L_2,L_3}=d_{L_1,L_2,L_3}+\delta_{L_1+L_2,L_3-2}+\delta_{L_3+L_1,L_2-2}-\delta_{L_2+L_3,L_1}\comma
\end{aligned}
\eeq
we can rewrite the three-point function also as
\beq
\begin{aligned}
\left.\langle  \NO{\tilde{\Phi}^{L_1}}\NO{\tilde{\Phi}^{L_2}}\NO{\tilde{\Phi}^{L_3}}\rangle\right|_{O(g^2)}=&\frac{(2\pi g)^2(-g^2)^{\frac{L_{\rm tot}}{2}}}{6} (3d_{L_1,L_2,L_3}-\delta_{L_1+L_2,L_3}\\
&\qquad\qquad\qquad\qquad -\delta_{L_2+L_3,L_1}-\delta_{L_3+L_1,L_2})\period
\end{aligned}
\eeq
By dividing the correlators by the expectation value of the Wilson loop $\langle \mathcal{W}\rangle=1+2\pi^2g^2+O(g^2)$, we get the following results for the normalized correlators:
\beq
\begin{aligned}
\left.\langle \!\langle \NO{\tilde{\Phi}^{L_1}}\NO{\tilde{\Phi}^{L_2}}\rangle\!\rangle\right|_{O(g^2)}=&-\frac{2(\pi g)^2(-g^2)^{L_1}}{3} \delta_{L_1,L_2}\comma\\
\left.\langle \!\langle \NO{\tilde{\Phi}^{L_1}}\NO{\tilde{\Phi}^{L_2}}\NO{\tilde{\Phi}^{L_3}}\rangle\!\rangle\right|_{O(g^2)}=&-\frac{(2\pi g)^2(-g^2)^{\frac{L_{\rm tot}}{2}}}{6} (\delta_{L_1+L_2,L_3}+\delta_{L_2+L_3,L_1}+\delta_{L_3+L_1,L_2})\period
\end{aligned}
\eeq
They are in perfect agreement with the direct one-loop computation performed in \cite{KK}.

For completeness, let us also present the structure constant in the standard CFT normalization; namely the normalization in which the two-point functions become unity. The result up to $O(g^2)$ reads
\begin{align}
&\frac{\langle\!\langle  \NO{\tilde{\Phi}^{L_1}}\NO{\tilde{\Phi}^{L_2}}\NO{\tilde{\Phi}^{L_3}}\rangle\!\rangle}{\left(\langle\!\langle  \NO{\tilde{\Phi}^{L_1}}\NO{\tilde{\Phi}^{L_1}}\rangle\!\rangle\langle\!\langle  \NO{\tilde{\Phi}^{L_2}}\NO{\tilde{\Phi}^{L_2}}\rangle\!\rangle\langle\!\langle  \NO{\tilde{\Phi}^{L_3}}\NO{\tilde{\Phi}^{L_3}}\rangle\!\rangle\right)^{1/2}}=\nn\\
&d_{L_1,L_2,L_3}+(\pi g)^2\left[d_{L_1,L_2,L_3}-\frac{2}{3}(\delta_{L_1+L_2,L_3}+\delta_{L_2+L_3,L_1}+\delta_{L_3+L_1,L_2})\right]\period
\end{align} 
\subsection{Strong coupling expansion}
Let us consider the expansion at strong coupling. Here we send $g\to \infty$ while keeping the lengths of the operators $L_i$'s finite. In this limit, the integral
\beq
\langle  \NO{\tilde{\Phi}^{L_1}}\NO{\tilde{\Phi}^{L_2}}\cdots\NO{\tilde{\Phi}^{L_n}} \rangle=\frac{2}{\pi}\int^{1}_{-1}dy \frac{\sinh (4\pi g \sqrt{1-y^2})}{4\pi g} \prod_{k=1}^{n}Q_{L_k}(x)\comma
\eeq
can be approximated by its saddle point,
\beq
\left.\frac{\del \log \sinh (4\pi g \sqrt{1-y^2})}{\del y}\right|_{y=y^{\ast}}=0\quad \Rightarrow \quad y^{\ast}=0\period
\eeq
Expanding the measure around this saddle point and performing the change of variables $t=\sqrt{2\pi g} y$, we obtain the following expression for the measure at strong coupling:
\beq\label{eq:strongmeasure}
\frac{2}{\pi}\int^{1}_{-1}dy\frac{\sinh (4\pi g\sqrt{1-y^2})}{4\pi g}=\int^{\infty}_{-\infty}\left(d\bar{\mu}^{0}(t)+\frac{1}{g}d\bar{\mu}^{1}(t)+O(g^2)\right)+O (e^{-g})\period
\eeq
with
\beq
\begin{aligned}\label{eq:strongmeasureexpansion}
d\bar{\mu}^{0}=\frac{e^{4\pi g}}{(2\pi)^{5/2}g^{3/2}}\,\, e^{-t^2}dt\comma\qquad d\bar{\mu}^{1}=d\bar{\mu}^{0}\times \left(-\frac{t^4}{8\pi}\right)\comma
\end{aligned}
\eeq
At the leading order, the measure $d\mu^{0}$ is simply a gaussian. As is well-known, this is nothing but the measure for the Hermite polynomials. Thus, $Q_L(x)$ at strong coupling is given by
\beq\label{eq:QN0strong}
\bar{Q}^{0}_L(x)=(-i)^{L}\left(\frac{g}{2\pi}\right)^{L/2}H_L (t)\period
\eeq 
Here the factor $(-i)^{L}(g/2\pi)^{L/2}$ comes from the normalization of $Q_L$ \eqref{eq:QLInormalization}, and $H_L(t)$ is the Hermite polynomial defined by
\beq
\begin{aligned}
&H_0(t)=1\comma\qquad H_1(t)=2t\comma\\
&H_L(t)=2t H_{L-1}(t)-2(L-1)H_{L-2}(t)\period
\end{aligned}
\eeq
We can then compute the two- and the three-point functions using the properties of the Hermite polynomials,
\beq
\begin{aligned}
&\int_{-\infty}^{\infty} dt\,e^{-t^2} H_L(t)H_M(t)=2^{L}L! \sqrt{\pi}\delta_{LM}\comma\\
&H_L(t)H_M(t)=\sum_{k=0}^{M}\frac{2^{M-k}L! M!}{(L-M+k)!(M-k)!k!}H_{L-M+2k}(t)\qquad (L\geq M)\period
\end{aligned}
\eeq
The results are given by
\beq
\begin{aligned}
&\left.\langle  \NO{\tilde{\Phi}^{L_1}}\NO{\tilde{\Phi}^{L_2}}\rangle\right|_{g\to \infty}=\frac{e^{4\pi g}}{2(2g)^{3/2}\pi^2}\left(-\frac{g}{\pi}\right)^{L_1}L_1!\delta_{L_1L_2}\comma\\
&\left.\langle  \NO{\tilde{\Phi}^{L_1}}\NO{\tilde{\Phi}^{L_2}}\NO{\tilde{\Phi}^{L_3}}\rangle\right|_{g\to \infty}\!\!\!=\frac{e^{4\pi g}}{2(2g)^{3/2}\pi^2}\left(-\frac{g}{\pi}\right)^{\frac{L_{\rm tot}}{2}}\frac{L_1!L_2!L_3!\,\,d_{L_1,L_2,L_3}}{L_{12|3}!
L_{23|1}!L_{31|2}!}\comma
\end{aligned}
\eeq
with $L_{ij|k}\equiv (L_i+L_j-L_k)/2$.
Note that the overall coefficient $e^{4\pi g}/(2(2g)^{3/2}\pi^2)$ is precisely the expectation value of the circular Wilson loop at strong coupling. Therefore, the normalized correlators take the following simple form:
\beq
\begin{aligned}
&\left.\langle\!\langle  \NO{\tilde{\Phi}^{L_1}}\NO{\tilde{\Phi}^{L_2}}\rangle\!\rangle\right|_{g\to \infty}=\left(-\frac{g}{\pi}\right)^{L_1}L_1!\delta_{L_1L_2}\comma\\
&\left.\langle\!\langle  \NO{\tilde{\Phi}^{L_1}}\NO{\tilde{\Phi}^{L_2}}\NO{\tilde{\Phi}^{L_3}}\rangle\!\rangle\right|_{g\to \infty}\!\!\!=\left(-\frac{g}{\pi}\right)^{\frac{L_{\rm tot}}{2}}\frac{L_1!L_2!L_3!\,\,d_{L_1,L_2,L_3}}{L_{12|3}!
L_{23|1}!L_{31|2}!}\period
\label{loc-leading}
\end{aligned}
\eeq
These results reproduce the strong-coupling answer, which is given by the generalized free fields in AdS$_2$.

Let us now compute the correction to this strong coupling answer. At the next order, the measure receives a correction $d\mu^{1}$, given by \eqref{eq:strongmeasureexpansion}.
As in the weak-coupling analysis, the change of the measure induces the correction to $Q_L$ since they have to satisfy the modified orthogonality condition:
\beq
\begin{aligned}
-\int^{\infty}_{-\infty} d\bar{\mu}^{1} \bar{Q}^{0}_L  \bar{Q}^{0}_M+\int^{\infty}_{-\infty} d\bar{\mu}^{0} \bar{Q}^{1}_L \bar{Q}^{0}_M+\int^{\infty}_{-\infty} d\bar{\mu}^{0} \bar{Q}^{0}_L  \bar{Q}^{1}_M
\propto \delta_{LM}\period
\end{aligned}
\eeq
To solve this condition, we use the following property of the Hermite polynomial:
\beq
\begin{aligned}
t^4 H_L(t)=&\frac{H_4(t)+12H_2(t)+12}{16}H_L(t)\\
=&\frac{1}{16}H_{L+4}+\frac{2L+3}{4}H_{L+2}+\frac{3(2L^2+2L+1)}{4}H_L\\
&+(2L-1)\frac{L!}{(L-2)!}H_{L-2}+\frac{L!}{(L-4)!}H_{L-4}\period
\end{aligned}
\eeq
We then get
\beq
\bar{Q}_{L}^{1}(x)=\frac{(-i)^{L}}{8\pi}\left(\frac{g}{ 2\pi}\right)^{L/2}\left[(2L-1)\frac{L!}{(L-2)!}H_{L-2}(t)+\frac{L!}{(L-4)!}H_{L-4}(t)\right]\period
\eeq
Using this result, we can compute the correction to the two-point function as
\beq
\begin{aligned}
\left.\langle  \NO{\tilde{\Phi}^{L_1}}\NO{\tilde{\Phi}^{L_2}}\rangle\right|_{O(1/g)}=&-\frac{e^{4\pi g}}{2(2g)^{3/2}\pi^2}\left(-\frac{g}{\pi}\right)^{L_1}L_1!\delta_{L_1L_2}\frac{3}{32\pi g}(2L_1^2+2L_1+1)\comma\\
\left.\langle  \NO{\tilde{\Phi}^{L_1}}\NO{\tilde{\Phi}^{L_2}}\NO{\tilde{\Phi}^{L_3}}\rangle\right|_{O(1/g)}\!\!\!=&-\frac{e^{4\pi g}}{2(2g)^{3/2}\pi^2}\left(-\frac{g}{\pi}\right)^{\frac{L_{\rm tot}}{2}}\frac{3}{64\pi g}(L_{\rm tot}^2+2L_{\rm tot}+2) \\
& \times\frac{L_1!L_2!L_3!\,\,d_{L_1,L_2,L_3}}{L_{12|3}!
L_{23|1}!L_{31|2}!} \period
\end{aligned}
\eeq
Since the expectation value of the Wilson loop can be expanded at strong coupling as
\beq
\langle\mathcal{W} \rangle\overset{g\to \infty}{=}\frac{e^{4\pi g}}{2(2g)^{3/2}\pi^2}\left(1-\frac{3}{32\pi g}+O(1/g^2)\right)\comma
\eeq
the normalized correlators are given by
\beq
\begin{aligned}
\left.\langle \!\langle \NO{\tilde{\Phi}^{L_1}}\NO{\tilde{\Phi}^{L_2}}\rangle\!\rangle\right|_{O(1/g)}=&-\left(-\frac{g}{\pi}\right)^{L_1}L_1!\delta_{L_1L_2}\frac{3}{32\pi g}(2L_1^2+2L_1)\comma\\
\left.\langle\!\langle  \NO{\tilde{\Phi}^{L_1}}\NO{\tilde{\Phi}^{L_2}}\NO{\tilde{\Phi}^{L_3}}\rangle\!\rangle\right|_{O(1/g)}\!\!\!=&-\left(-\frac{g}{\pi}\right)^{\frac{L_{\rm tot}}{2}}\frac{3}{64\pi g}(L_{\rm tot}^2+2L_{\rm tot})  \frac{L_1!L_2!L_3!\,\,d_{L_1,L_2,L_3}}{L_{12|3}!
L_{23|1}!L_{31|2}!} \period
\label{normalized}
\end{aligned}
\eeq
As we will see in the next subsection, these results are in perfect agreement with the direct strong-coupling computation.

Using these results, we can also compute the structure constant in the standard CFT normalization at strong coupling:
\begin{align}
&\frac{\langle\!\langle  \NO{\tilde{\Phi}^{L_1}}\NO{\tilde{\Phi}^{L_2}}\NO{\tilde{\Phi}^{L_3}}\rangle\!\rangle}{\left(\langle\!\langle  \NO{\tilde{\Phi}^{L_1}}\NO{\tilde{\Phi}^{L_1}}\rangle\!\rangle\langle\!\langle  \NO{\tilde{\Phi}^{L_2}}\NO{\tilde{\Phi}^{L_2}}\rangle\!\rangle\langle\!\langle  \NO{\tilde{\Phi}^{L_3}}\NO{\tilde{\Phi}^{L_3}}\rangle\!\rangle\right)^{1/2}}=\nn\\
&\frac{\sqrt{L_1!L_2!L_3!}\,\,d_{L_1,L_2,L_3}}{L_{12|3}!
L_{23|1}!L_{31|2}!} \left[1+\frac{3(L_1^2+L_2^2+L_3^2-2(L_1L_2+L_2L_3+L_3L_1))}{64\pi g} +O\left(\frac{1}{g^2}\right)\right] \period
\end{align}

\subsection{Comparison to string theory}
In this section we show that the strong coupling expansion of the localization results derived above precisely matches the direct 
perturbative calculation using the AdS$_5 \times S^5$ string sigma model. As is well-known, on the string theory side the 1/2-BPS (circular or straight) 
Wilson loop is dual to a minimal surface with the geometry of an AdS$_2$ embedded in AdS$_5$ (and pointlike in the $S^5$ directions). 
The dynamics 
of the string worldsheet fluctuations is most conveniently described using the Nambu-Goto action in static gauge. The bosonic 
part of the string action up to the quartic order was written down explicitly in \cite{GRT} and it reads
\def \del{\partial}
\begin{equation}
S_B = \frac{\sqrt{\lambda}}{2\pi}\int d^2\sigma \sqrt{g}\left(
1+\frac{1}{2}  g^{\mu\nu}\del_{\mu} x^i \del_{\nu} x^i  +  x^i x^i 
+ \frac{1}{2} g^{\mu\nu}\del_{\mu} y^a \del_{\nu} y^a + L_{4y}+L_{4x}+L_{2x,2y}+\ldots \right)
\end{equation}
Here $g_{\mu\nu}$ is the AdS$_2$ worldsheet metric, $y^a, a=1,\ldots, 5$ are the massless fluctuations in the $S^5$ directions, which are dual to the scalar insertions $\Phi^a$ on the gauge theory side, 
and $x^i, i=1,2,3$ are the the $m^2=2$ fluctuations in AdS$_5$ dual to insertions of the displacement operator \cite{Yoshida,Yoshida2}. For the explicit form of the quartic vertices, see \cite{GRT}. Note that there are no cubic vertices between the elementary bosonic 
fluctuations. 

Let us first review the result for the tree-level connected four-point function of the $y^a$ fluctuations computed in \cite{GRT}, and its 
agreement with the localization prediction. Taking the circular 
geometry at the boundary, it takes the form
\begin{equation}
\begin{aligned}
&\langle Y_1 \cdot y(\tau_1) Y_2 \cdot y(\tau_2)Y_3 \cdot y(\tau_3)Y_4 \cdot y(\tau_4) \rangle_{{\rm AdS}_2}^{\rm conn.} = \\
&=\frac{\left(\frac{\sqrt{\lambda}}{2\pi^2}\right)^2 Y_1\cdot Y_2 \,Y_3\cdot Y_4}{(4\sin\frac{\tau_{12}}{2}\sin\frac{\tau_{34}}{2})^2}
\frac{1}{\sqrt{\lambda}}\left[ G_S(\chi)-\frac{2}{5}G_T(\chi)
+ \xi (G_T(\chi)+G_A(\chi))+ \zeta (G_T(\chi)-G_A(\chi))\right]\,.
\label{4pt-connected}
\end{aligned}
\end{equation}
where $\chi$ is the cross-ratio
\begin{equation}
\chi = \frac{\sin\frac{\tau_{12}}{2}\sin\frac{\tau_{34}}{2}}{\sin\frac{\tau_{13}}{2}\sin\frac{\tau_{24}}{2}}
\end{equation}
and $\xi,\zeta$ are $SO(5)$ cross-ratios
\begin{equation}
\xi = \frac{Y_1\cdot Y_3\, Y_2\cdot Y_4}{Y_1\cdot Y_2 \,Y_3\cdot Y_4}\,,\qquad 
\zeta = \frac{Y_1\cdot Y_4 \,Y_2\cdot Y_3}{Y_1\cdot Y_2 \,Y_3\cdot Y_4}\,,
\end{equation}
with $Y_i$ null polarization 5-vectors. The functions of cross-ratio $G_{S,T,A}(\chi)$ appearing in the 4-point function above correspond to singlet, 
symmetric traceless and antisymmetric channels, and their explicit form can be found in \cite{GRT}. 

In writing (\ref{4pt-conn}) we have taken the normalization of the $y$ fluctuations such that the leading 
order 2-point function computed from the string action reads\footnote{In \cite{GRT} instead a canonical normalization of the kinetic term was used, 
so that the leading 2-point function was $\lambda$ independent. The normalization in (\ref{2pt-norm}) is actually the one which is naturally induced 
by the overall $\lambda$ dependence in the string action, upon adopting the standard AdS/CFT dictionary to compute the tree-level 2-point function, see 
\cite{Fiol:2013iaa}.} 
\begin{equation}
\langle Y_1 \cdot y(\tau_1) Y_2\cdot y(\tau_2) \rangle_{{\rm AdS}_2} =\frac{\sqrt{\lambda}}{2\pi^2} \frac{Y_1\cdot Y_2}{(2\sin\frac{\tau_{12}}{2})^2}\,. 
\label{2pt-norm}
\end{equation}
This normalization agrees in the strong coupling limit with the normalization we adopted on the gauge theory side, which gives 
\begin{equation}
\begin{aligned}
\langle\! \langle Y_1 \cdot \Phi(\tau_1)Y_2 \cdot \Phi(\tau_2) \rangle \!\rangle 
= \frac{\sqrt{\lambda } \, I_2(\sqrt{\lambda })}{2\pi ^2\,  I_1(\sqrt{\lambda })}\frac{Y_1\cdot Y_2}{(2\sin\frac{\tau_{12}}{2})^2}
=\frac{\sqrt{\lambda }}{2\pi^2}\left(1-\frac{3}{2\sqrt{\lambda}}+\ldots \right)\frac{Y_1\cdot Y_2}{(2\sin\frac{\tau_{12}}{2})^2}\,.
\end{aligned}
\label{2pt-exact}
\end{equation}  
We now specialize to the topological boundary operators, by choosing the polarizations 
\begin{equation}
Y_i = (\cos\tau_i,\sin\tau_i,0,i,0,0)\,.
\end{equation} 
By analogy with the notation introduced earlier on the CFT side, let us define
\begin{equation}
\tilde y(\tau) \equiv \cos(\tau)y^1(\tau)+\sin(\tau)y^2(\tau)+i y^4(\tau) \,,
\end{equation}
which is dual to the insertion of $\tilde\Phi$ and has the constant 2-point function given at leading order by 
$
\langle \tilde y(\tau_1)\tilde y(\tau_2)\rangle_{{\rm AdS}_2}  = -\frac{\sqrt{\lambda}}{4\pi^2}
$. Then, using the explicit form of $G_{S,T,A}(\chi)$, one finds the position independent result for the connected 
4-point function
\begin{equation}
\langle \tilde y(\tau_1) \tilde y(\tau_2) \tilde y(\tau_3) \tilde  y(\tau_4) \rangle_{{\rm AdS}_2}^{\rm conn.}  
=-\frac{3 \sqrt{\lambda }}{16 \pi ^4}\,.
\label{4pt-conn}
\end{equation}
\begin{figure}
\centering
\includegraphics[clip,height=7cm]{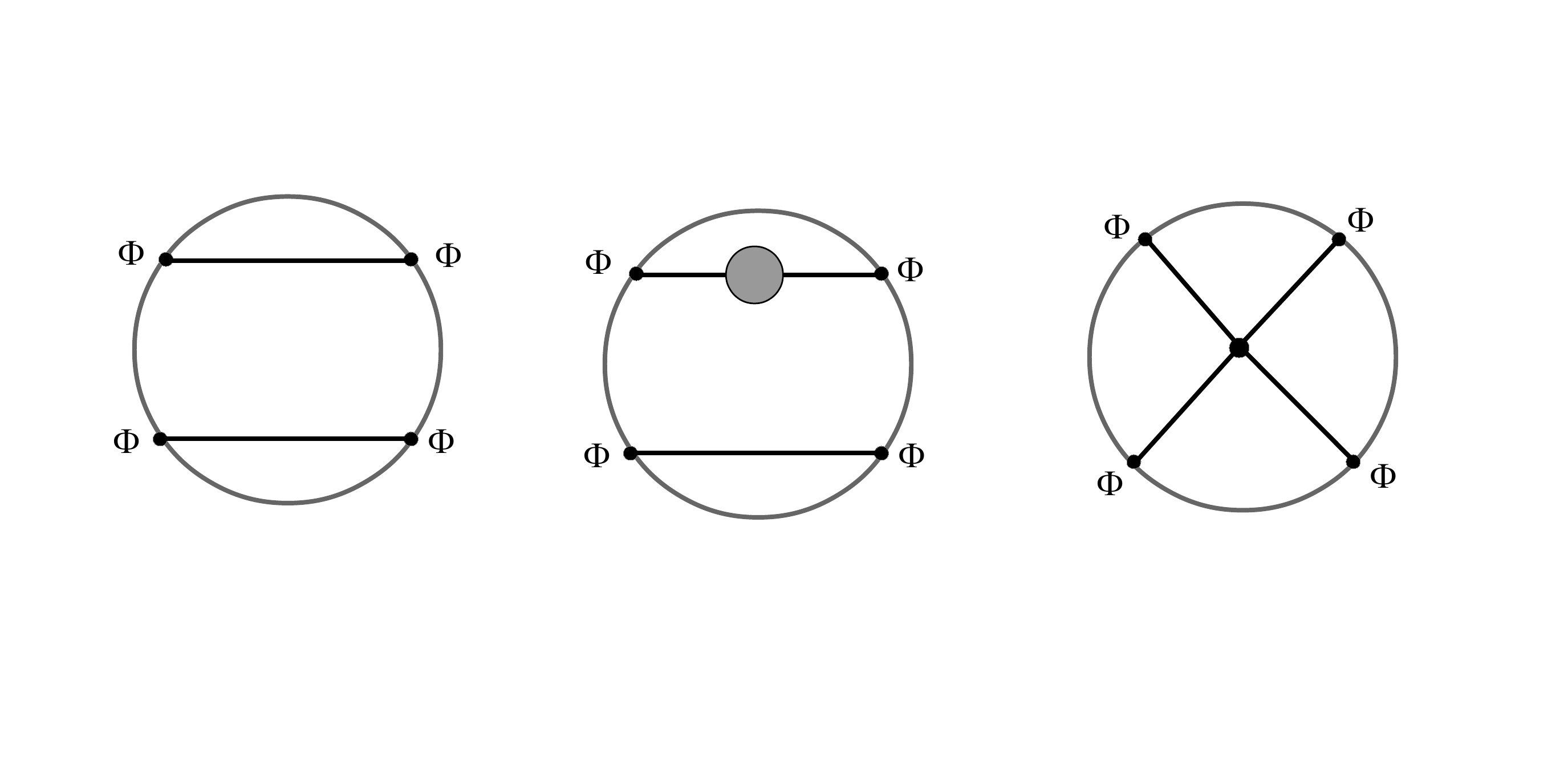}
\vskip -1.3cm
\caption{Witten diagrams in AdS$_2$ contributing to the 4-point function of single-letter insertions $\Phi$ to next-to-leading order at strong coupling. The 
grey blob in the middle figure denote the one-loop correction to the ``boundary-to-boundary'' $y$ propagator.}
\label{fig:4pt-y}
\end{figure}
The full 4-point function to the first subleading order also receives contribution from disconnected diagrams, as shown in 
figure \ref{fig:4pt-y}. In addition to the leading tree-level generalized free-field Wick contractions, there are corrections of the same order as 
(\ref{4pt-conn}) coming from disconnected diagrams where one leg is one-loop corrected, see the figure. While these corrections have not been computed 
explicitly yet from string theory, we will assume below that they reproduce the strong coupling expansion of (\ref{2pt-exact}).\footnote{Alternatively, one may consider 
normalized correlators as in (\ref{normalized}), where such corrections drop out in the ratio.} Then, the 4-point function of single-letter insertions 
computed from the AdS$_2$ string theory side reads to this order 
\begin{equation}
\langle \tilde y(\tau_1) \tilde y(\tau_2) \tilde y(\tau_3) \tilde  y(\tau_4) \rangle_{{\rm AdS}_2} 
= \left(-\frac{\sqrt{\lambda}}{4\pi^2}\right)^2\left[\left(1-\frac{3}{2\sqrt{\lambda}}\right)^2-\frac{3}{\sqrt{\lambda}}+\ldots \right]
= \frac{3 \lambda }{16 \pi ^4}-\frac{3 \sqrt{\lambda }}{4 \pi ^4}+\ldots 
\end{equation}
where the first term in the bracket is the contribution of disconnected diagrams, and the second term the one of the tree-level connected diagram. 
This precisely matches the strong coupling expansion of the localization result 
\begin{eqnarray}
\langle \!\langle \tilde \Phi \tilde \Phi \tilde \Phi \tilde \Phi\rangle\! \rangle &=& 
\frac{\frac{\del^4}{\del A^4} \langle  {\cal W} \rangle}{\langle {\cal W}\rangle}\Big{|}_{A=2\pi} =
\frac{3 \lambda }{16 \pi ^4}+\frac{3}{2 \pi ^4}-\frac{3 \sqrt{\lambda } I_0(\sqrt{\lambda })}{4 \pi ^4 I_1(\sqrt{\lambda })}\,.
\end{eqnarray}

\begin{figure}
\centering
\includegraphics[clip,height=7cm]{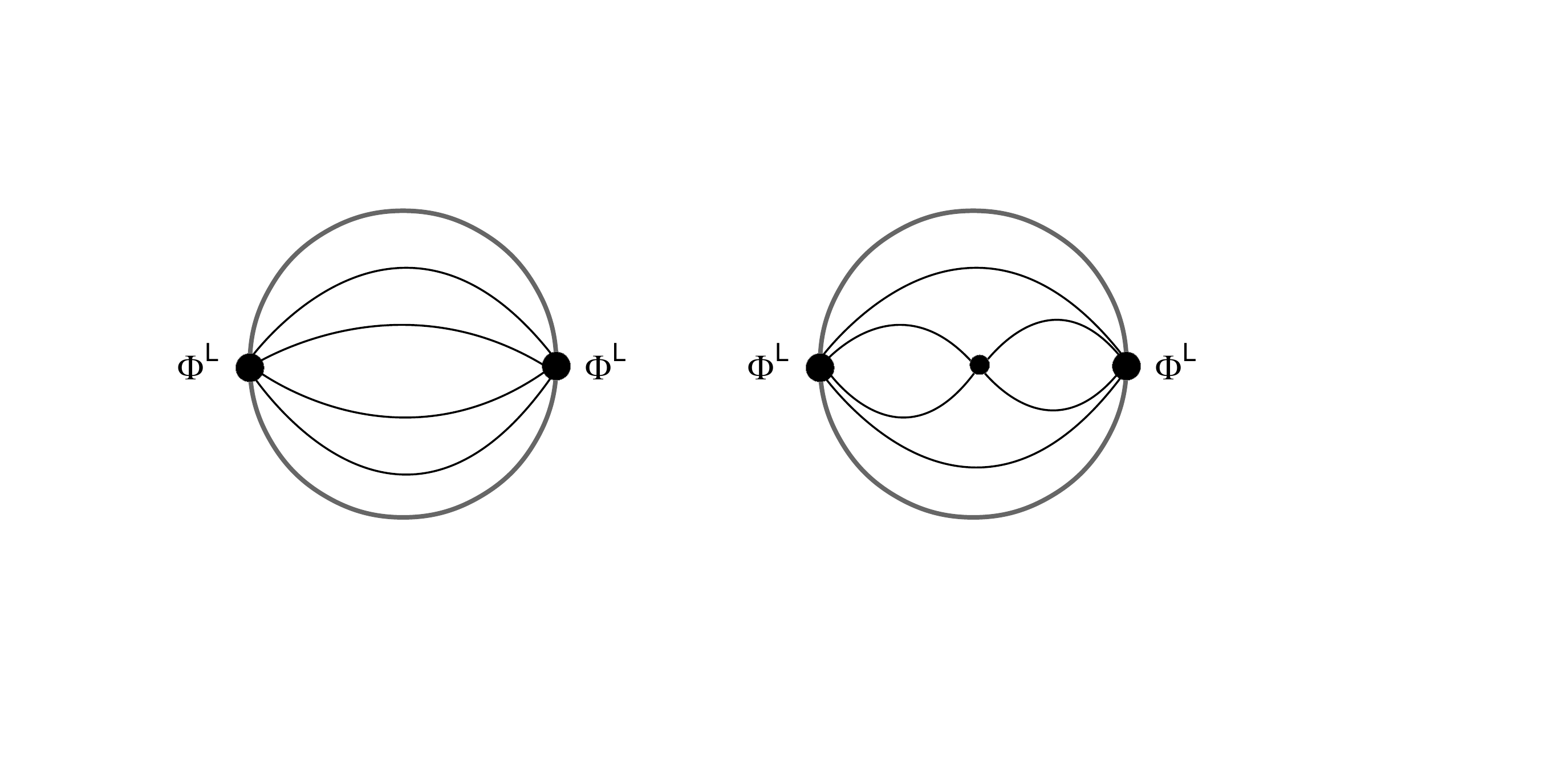}
\vskip -1.3cm
\caption{Topology of Witten diagrams contributing to the 2-point function of $\Phi^L$ (in the picture the case $L=4$ is shown). 
The diagrams on the left, corresponding to generalized free-field contractions, 
also receive a subleading correction where a $y$-propagator is one-loop corrected.}
\label{fig:2pt}
\end{figure}

Having reviewed the matching of the $\tilde \Phi$ 4-point function, let us now move to the computation of the two-point and three-point functions 
of arbitrary length insertions $\NO{\tilde \Phi^L}$. The Witten diagrams contributing to the 2-point function to the first two orders in the strong coupling 
expansion are given in figure \ref{fig:2pt} (as in figure \ref{fig:4pt-y} above, there are one-loop corrections to the diagrams involving free-field Wick 
contractions, that for brevity 
we do not depict in the figure). The contribution of the diagram involving the 4-point vertex can be obtained from the 4-point result (\ref{4pt-connected}) by 
taking $Y_2\rightarrow Y_1, Y_{3,4}\rightarrow Y_2$, and taking the limit $\tau_2\rightarrow \tau_1, \tau_4\rightarrow \tau_3 \equiv \tau_2$. 
From \cite{GRT}, we have $G_T(\chi)=-\frac{3}{2}\chi^2+\ldots$ and $G_A(\chi)=O(\chi^3\log(\chi))$ at small $\chi$, and so we get  
\begin{equation}
\langle (Y_1\cdot y(\tau_1))^2 (Y_2\cdot y(\tau_2))^2 \rangle_{{\rm AdS}_2} =  
-\frac{3 \sqrt{\lambda} (Y_1\cdot Y_2)^2}{64 \pi ^4\sin^4\frac{\tau_{12}}{2}} = -\frac{3 \sqrt{\lambda }}{16 \pi ^4}\,,
\label{2pt-limit}
\end{equation} 
where the first equality is valid for any choice of the null polarization vectors, and in the second equality we have specialized to the topological 
configuration. We can now use this result and some elementary combinatorics to compute the 2-point functions for arbitrary length. We find
\begin{eqnarray}
\langle \tilde y^L \tilde y^L \rangle_{{\rm AdS}_2} = \left(-\frac{\sqrt{\lambda}}{4\pi^2}\right)^L
\left[L! \left(1-\frac{3L}{2\sqrt{\lambda}}\right)-\frac{3}{\sqrt{\lambda}} \begin{pmatrix} L \\ 2 \end{pmatrix}^2 (L-2)! +\ldots \right] 
\end{eqnarray} 
The first term in brackets corresponds to the generalized free field Wick contractions: there are clearly $L!$ such contractions, and the 
factor $(1-\frac{3}{2\sqrt{\lambda}}+\ldots)^L=1-\frac{3L}{2\sqrt{\lambda}}+\ldots$ accounts for the one-loop correction of the boundary-to-boundary legs, as discussed 
above. The second term in brackets corresponds to the diagrams involving the 4-point vertex shown in figure \ref{fig:2pt}: 
there are $\begin{pmatrix} L \\ 2 \end{pmatrix}$ ways of picking two 
$y$'s on each operator, and $(L-2)!$ free-field contractions among the remaining $y$'s. This result can be simplified to
\begin{equation}
\langle \tilde y^L \tilde y^L \rangle_{{\rm AdS}_2} 
 = \left(-\frac{\sqrt{\lambda}}{4\pi^2}\right)^L\, L! \left[1-\frac{3}{4\sqrt{\lambda}}L(L+1)+\ldots \right]\,,
\end{equation}
which indeed precisely agrees with the localization result given in (\ref{loc-leading}) and (\ref{normalized}).

\begin{figure}
\centering
\includegraphics[clip,height=7cm]{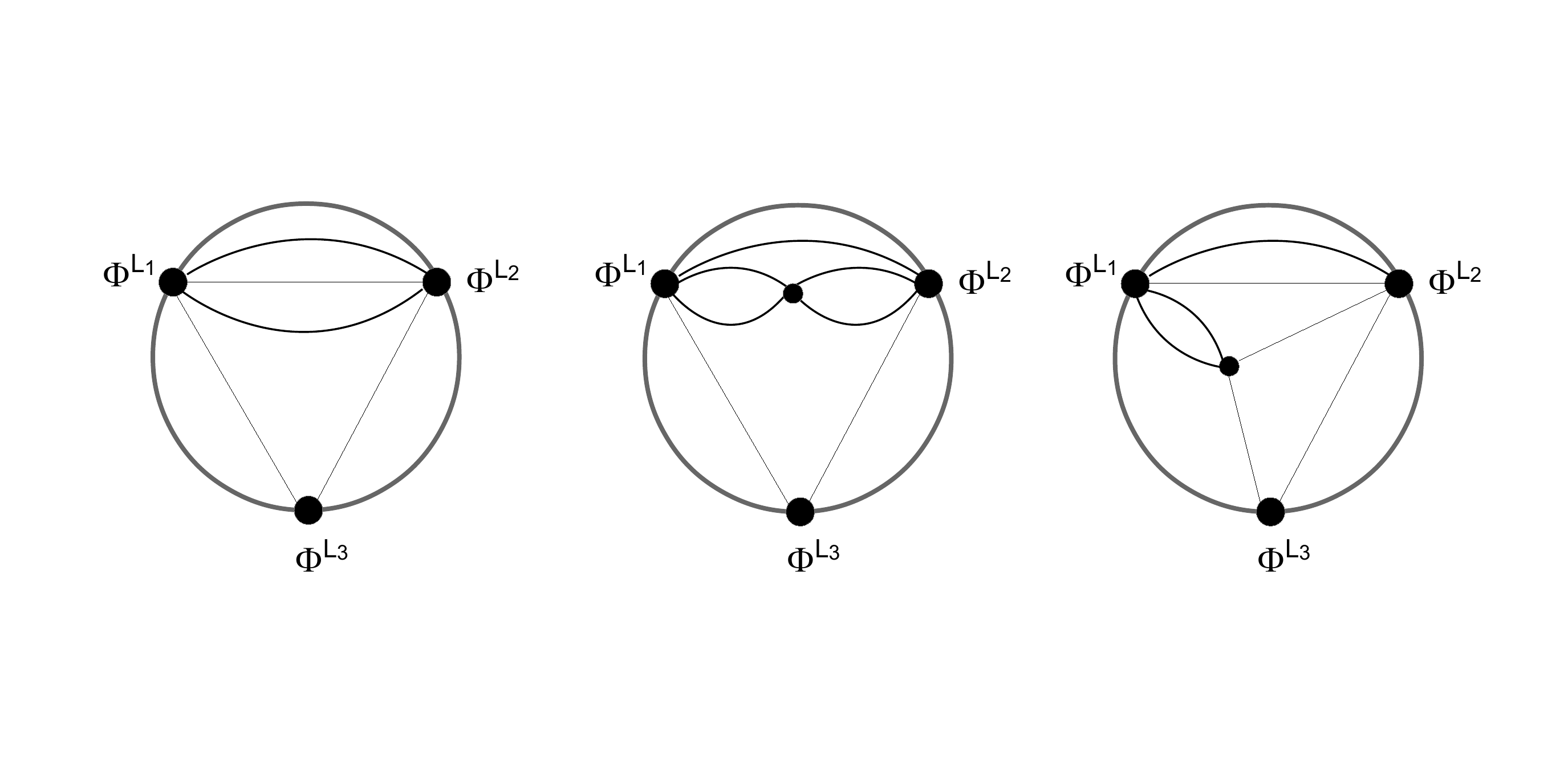}
\vskip -1.3cm
\caption{Topology of Witten diagrams contributing to the 3-point function of general length operators 
(in the picture the case $L_1=L_2=4, L_3=2$ is shown). In addition, there are one-loop corrections to the generalized free-field diagrams shown on 
the left.}
\label{fig:3pt}
\end{figure}

Similarly, the diagrams contributing to the 3-point function $\langle\! \langle  \NO{\tilde\Phi^{L_1}}\NO{\tilde \Phi^{L_2}}\NO{\tilde \Phi^{L_3}}\rangle \!\rangle$ 
are shown in figure \ref{fig:3pt}. The leading contribution is given again by free-field Wick contractions. Let us define the number of such 
contractions to be 
\begin{equation}
{\rm n}_{L_1,L_2,L_3} \equiv \frac{L_1!L_2!L_3!\,\,d_{L_1,L_2,L_3}}{L_{12|3}!
L_{23|1}!L_{31|2}!}
\end{equation}
with $d_{L_1,L_2,L_3}$ given in (\ref{eq:defofdfunctions}). At the subleading order, there are two topologies which involve the 4-point vertex: one 
where the vertex connects two $y$'s belonging to two different operators, and one where it connects two $y$'s from one operator and two $y$'s from two 
separate operators, see the figure.\footnote{Note that there is no diagram where the 4-vertex connects three $y$'s on the same operator, as this vanishes 
by $SO(5)$ symmetry: in terms of the null polarization vectors, it necessarily involves a factor $Y_i \cdot Y_i = 0$.} The first type of diagram 
can be computed using (\ref{2pt-limit}). For the second type of diagram, again taking the limit of the 4-point result (\ref{4pt-conn}) by 
setting $Y_2\rightarrow Y_1, Y_3\equiv Y_2, Y_4\equiv Y_3$ and similarly for the $\tau_i$ points, one finds
\begin{equation}
\langle (Y_1\cdot y(\tau_1))^2 \, Y_2\cdot y(\tau_2)  \, Y_3\cdot y(\tau_3)\rangle_{{\rm AdS}_2} =  
-\frac{3 \sqrt{\lambda} (Y_1\cdot Y_2) (Y_1\cdot Y_3)}{64 \pi ^4\sin^2\frac{\tau_{12}}{2}\sin^2\frac{\tau_{13}}{2}} 
= -\frac{3 \sqrt{\lambda }}{16 \pi ^4}\,,
\label{3pt-limit}
\end{equation}
where we have specialized to the topological configuration in the second step, but the first equality holds in general. 
Then, working out the relevant combinatorics and putting all the contributions together, we find for general lengths
\begin{equation}
\begin{aligned}
&\langle \tilde y^{L_1}\tilde y^{L_2}\tilde y^{L_3} \rangle_{{\rm AdS}_2}
= \left(-\frac{\sqrt{\lambda}}{4\pi^2}\right)^{\frac{L_{\rm tot}}{2}}\Big{[} 
{\rm n}_{L_1,L_2,L_3}(1-\frac{3(L_1+L_2+L_3)}{2\sqrt{\lambda}}+\ldots) \\
&
-\frac{3}{\sqrt{\lambda}}\left(
{\rm n}_{L_1-2,L_2-2,L_3} \begin{pmatrix} L_1\\2 \end{pmatrix} \begin{pmatrix} L_2\\2 \end{pmatrix}
+{\rm n}_{L_1-2,L_2,L_3-2} \begin{pmatrix} L_1\\2 \end{pmatrix} \begin{pmatrix} L_3\\2 \end{pmatrix}
+{\rm n}_{L_1,L_2-2,L_3-2} \begin{pmatrix} L_2\\2 \end{pmatrix} \begin{pmatrix} L_3\\2 \end{pmatrix}
\right)\cr 
& -\frac{3}{\sqrt{\lambda}}\left(
{\rm n}_{L_1-2,L_2-1,L_3-1} \begin{pmatrix} L_1\\2 \end{pmatrix} L_2 L_3
+{\rm n}_{L_1-1,L_2-2,L_3-1} \begin{pmatrix} L_2\\2 \end{pmatrix} L_1 L_3
+{\rm n}_{L_1-1,L_2-1,L_3-2} \begin{pmatrix} L_3\\2 \end{pmatrix} L_1 L_2
\right)\Big{]} \,,
\end{aligned}
\end{equation}
with $L_{\rm tot}=L_1+L_2+L_3$. This simplifies to
\begin{equation}
\langle \tilde y^{L_1}\tilde y^{L_2}\tilde y^{L_3} \rangle_{{\rm AdS}_2} = 
\left(-\frac{\sqrt{\lambda}}{4\pi^2}\right)^{\frac{L_{\rm tot}}{2}}{\rm n}_{L_1,L_2,L_3}
\left[1-\frac{3 L_{\rm tot} (L_{\rm tot}+2)}{16 \sqrt{\lambda }}+\ldots \right]\,,
\end{equation}
again in complete agreement with the localization prediction (\ref{loc-leading}) and (\ref{normalized}).

In a similar way, one can compute higher-point correlation functions of $\NO{\tilde \Phi^L}$ insertions to next-to-leading order at strong coupling. While 
for the topological operators the agreement of these should follow from the agreement of 2-point and 3-point functions shown above, 
to dispel any doubt we have explicitly verified in 
various higher-point examples that the localization results are indeed correctly reproduced by string perturbation theory around the AdS$_2$ minimal surface.

\section{Emergent matrix model at large $N$\label{sec:matrixmodel}}
In this section, we reformulate our results in the planar limit as a matrix model. We follow closely the approach in the integrability literature \cite{GS,SV}, but the resulting matrix model is slightly different. This reformulation would be useful for studying the semi-classical limit where $L_i$ and $g$ are both send to infinity while their ratios are kept finite. We present preliminary results for the semi-classical limit leaving more detailed analysis for future investigation.
\subsection{$D_L$ and $Q_L(x)$ as a matrix model}
Using the integral representations \eqref{eq:integralW} and \eqref{eq:integralWder}, the equation \eqref{eq:halfphiD} can be re-expressed as
\beq\label{eq:DLmeasuresym}
D_L=\left(\prod_{k=1}^{L}\oint d\mu_{\rm exp}(x_k)\right)\dmatrix{cccc}{1&X_1&\cdots&X_1^{L-1}\\X_2&X_2^2&\cdots&X_2^{L}\\\vdots&\vdots&\ddots&\vdots\\X_{L}^{L-1}&X_{L}^{L}&\cdots&X_{L}^{2L-2}}\comma
\eeq
with $X_i\equiv g(x_i-x_i^{-1})$. Here we used the exponential measure \eqref{eq:expmeasure} for later convenience, but the results in this subsection are equally valid if we substitute it with $d\mu$ or $d\mu_{\rm sym}$.
The determinant in \eqref{eq:DLmeasuresym} has the structure of the Vandermonde determinant and it can be rewritten as
\beq
\begin{aligned}\label{eq:rewritingstep0}
\dmatrix{cccc}{1&X_1&\cdots&X_1^{L-1}\\X_2&X_2^2&\cdots&X_2^{L}\\\vdots&\vdots&\ddots&\vdots\\X_{L}^{L-1}&X_{L}^{L}&\cdots&X_{L}^{2L-2}}&=\prod_k X_k^{k-1}\prod_{i<j}(X_j-X_i)\period
\end{aligned}
\eeq
Since the measure factors in \eqref{eq:DLmeasuresym} are symmetric under the permutation of the indices, we can replace the right hand side of \eqref{eq:rewritingstep0} with its symmetrized version,
\beq
\begin{aligned}\label{eq:rewritingstep1}
\prod_k X_k^{k-1}\prod_{i<j}(X_j-X_i) \to \frac{1}{L!}\prod_{i<j}(X_j-X_i)\sum_{\sigma\in S_{L}} (-1)^{|\sigma|}\prod_k X_{\sigma_k}^{k-1}\period
\end{aligned}
\eeq
We then realize that the sum over the permutation is precisely the definition of the Vandermonde determinant. 
We can thus replace the determinant part by
\beq
\dmatrix{cccc}{1&X_1&\cdots&X_1^{L-1}\\X_2&X_2^2&\cdots&X_2^{L}\\\vdots&\vdots&\ddots&\vdots\\X_{L}^{L-1}&X_{L}^{L}&\cdots&X_{L}^{2L-2}}\to \frac{\prod_{i<j}(X_i-X_j)^2}{L!}
\eeq

Therefore, we obtain the multi-integral expression,
\beq\label{eq:DLmatrix}
D_L =\frac{g^{L(L-1)}}{L!}\left(\prod_{k=1}^{L}\oint d\mu_{\rm exp}(x_k)\right) \prod_{i<j}(x_i-x_j)^2 \left(1+\frac{1}{x_i x_j}\right)^2\comma
\eeq
Note that this matrix-model-like expression is similar but different from the matrix model for $m_{2L}$, derived in \cite{SV}. One notable difference is that the integral in \cite{SV} contains $2L$ integration variables while the integral derived here contains only $L$ integration variables. As proven in section \ref{subsec:comparison}, the two determinants are related by \eqref{eq:relationmMdM}. 

One can also express the polynomial $F_L$ as a multiple integral. Applying the integral expression \eqref{eq:integralWder} to \eqref{eq:FnXexplicit}, we get
\beq
F_L[X]=\frac{1}{D_{L}}\left(\prod_{k=1}^{L}\oint d\mu_{\rm exp} (x_k)\right)\dmatrix{cccc}{1&X_1&\cdots&X_1^L\\X_2&X_2^2&\cdots&X_2^{L+1}\\\vdots&\vdots&\ddots&\vdots\\X_{L}^{L-1}&X_{L}^{L}&\cdots&X_{L}^{2L-1}\\1&X&\cdots &X^{L}}\period
\eeq
Here, again, $X_k\equiv x_k-x_k^{-1}$. 
Now the determinant part in the integrand can be evaluated as
\beq
\dmatrix{cccc}{1&X_1&\cdots&X_1^L\\X_2&X_2^2&\cdots&X_2^{L+1}\\\vdots&\vdots&\ddots&\vdots\\X_{L}^{L-1}&X_{L}^{L}&\cdots&X_{L}^{2L-1}\\1&X&\cdots &X^{L}}=\prod_k X_k^{k-1} (X-X_k)\prod_{i<j}(X_j-X_i)\period
\eeq
Thus, after symmetrization, we get
\beq
F_L(X)=\frac{g^{L(L-1)}}{L!D_{L}}\left[\prod_{k=1}^{L}\oint d\mu_{\rm exp}(x_k) \left(X- g(x_k-x_k^{-1})\right)\right]\prod_{i<j}(x_i-x_j)^2 \left(1+\frac{1}{x_i x_j}\right)^2\period
\eeq
As can be seen from this expression, $F_L(X)$ is the analogue of the characteristic polynomial of the matrix model, which is obtained by inserting $\det (X-M)$ in the integral of the matrix $M$.
After the change of the variables $X = g(x-x^{-1})$, it can be rewritten as
\beq
\begin{aligned}\label{eq:qlintegralexpression}
&Q_L(x)=F_L(g(x-x^{-1})) \\
&=\frac{g^{L^2}}{L!D_{L}}\left[\prod_{k=1}^{L}\oint d\mu_{\rm exp}(x_k) \,(x-x_k)\left(1+\frac{1}{x x_k}\right)\right]\prod_{i<j}(x_i-x_j)^2 \left(1+\frac{1}{x_i x_j}\right)^2\period
\end{aligned}
\eeq
\subsection{Classical limit of the matrix model}
Let us now consider the limit where $g$ and $L_i$'s are sent to infinity while their ratios remain finite. This limit corresponds to a classical string configuration in AdS and therefore is called the (semi-)classical limit.

The integral expression \eqref{eq:DLmatrix}  can be rewritten as
\beq
D_L =\frac{g^{L(L-1)}}{(4\pi g)^{L}L!}\oint \prod_{k=1}^{L}\frac{dx_k(1+x_k^{-2})}{2\pi i} \,e^{S_L(x_1,\ldots ,x_k)}\comma
\eeq 
where the action is given by
\beq
\begin{aligned}
S_L=&-\sum_{k=1}^{L}2\pi g\left(x_k+\frac{1}{x_k}\right)+ag \left(x_k-\frac{1}{x_k}\right)+2\sum_{i<j}^{L}\log \left[(x_i-x_j)\left(1+\frac{1}{x_ix_j}\right)\right]\period
\end{aligned}
\eeq
In the classical limit, the integral can be approximated by the saddle point $\del S_L/\del x_k=0$. To compare with the result from integrability, it is convenient to introduce the {\it rapidity} variables
\beq
u_k \equiv g \left(x_k +\frac{1}{x_k}\right)\period
\eeq
and express the saddle-point equation as $\del S_L/\del u_k =0$. We then get
\beq
\begin{aligned}\label{eq:saddlepoint}
\frac{1+x_k^2}{1-x_k^2}\left[\frac{a}{2}+\frac{1}{g}\sum^{L}_{j\neq k}\frac{1}{(x_k-x_j)\left(1+\frac{1}{x_k x_j}\right)}\right]=\pi\period
\end{aligned}
\eeq

\begin{figure}
\centering
\includegraphics[clip,height=7cm]{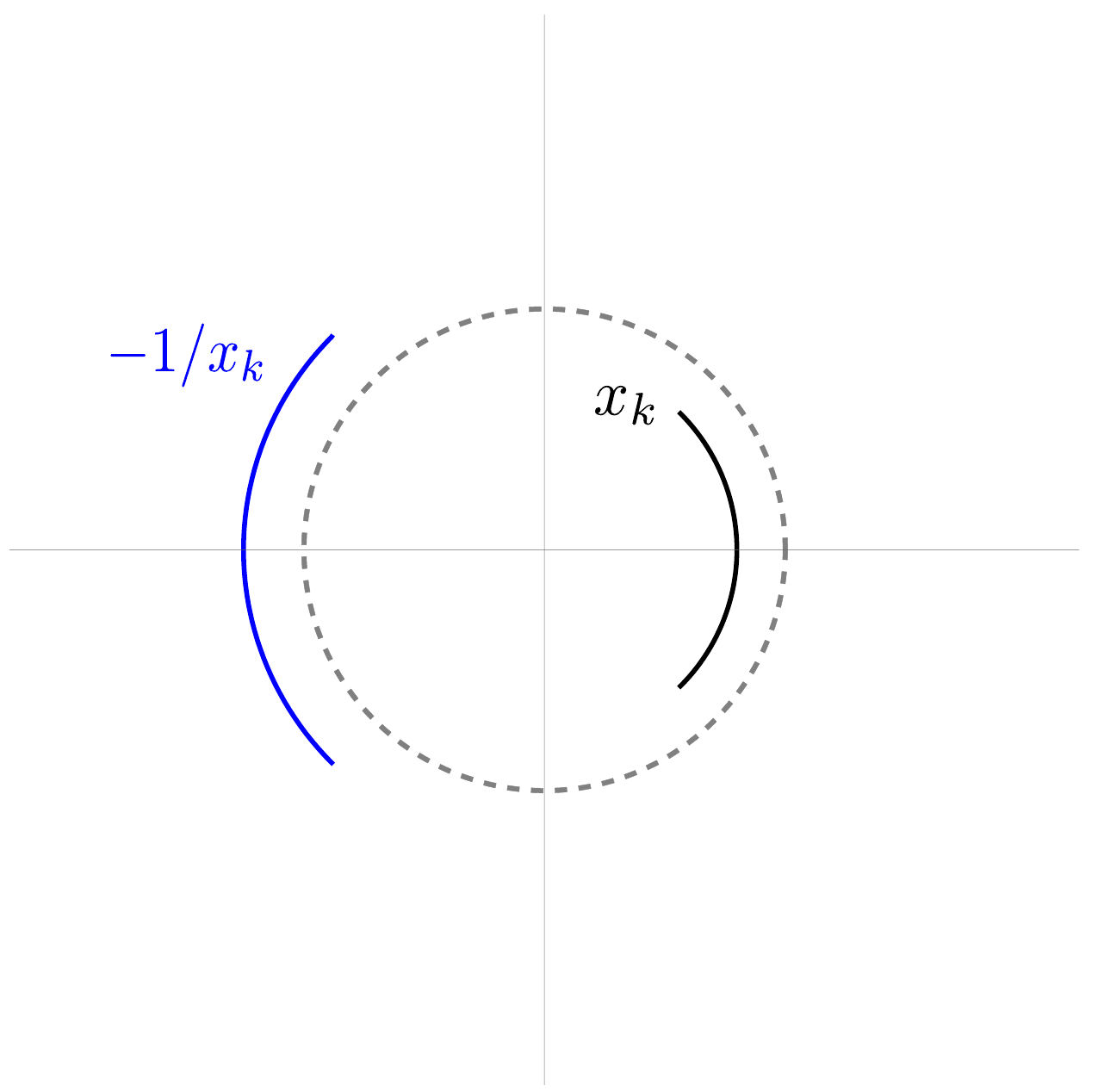}
\caption{Analytic structure of $p_L(x)$. In the semi-classical limit, the function $p_L(x)$ has two branch cuts; the one coming from the condensation of $x_k$ and the other coming from the condensation of $1/x_k$.}
\label{fig:branch}
\end{figure}
As in the usual large $N$ matrix models, we expect that $x_k$'s condense into a branch cut in the classical limit as shown in figure \ref{fig:branch}. To describe the limit, it is convenient to introduce a function $p_L(x)$ defined by
\beq\label{eq:defofquasimomentum}
p_L(x)\equiv \frac{1+x^2}{1-x^2}\left[\frac{a}{2g}+\frac{1}{g}\sum^{L}_{k=1}\frac{1}{(x-x_k)\left(1+\frac{1}{x x_k}\right)}\right]\period
\eeq 
Then, the saddle-point equation \eqref{eq:saddlepoint} can be rewritten as
\beq\label{eq:pxsaddlepoint}
\frac{1}{2}\left[p_L(x_k+\epsilon)+p_L(x_k-\epsilon)\right]=\pi\comma
\eeq
where $x_k\pm \epsilon$ denote the two different sides of the branch cut. Since $p(x)$ has the symmetry 
\beq\label{eq:sympx}
p_L(x)=-p_L(-1/x)\comma
\eeq
the branch cut of $x_k$'s is accompanied by another branch cut that is formed by $-1/x_k$'s. Around this other branch cut, $p_L(x)$ satisfies
\beq\label{eq:pxsaddlepoint2}
\frac{1}{2}\left[p_L\left(-x_k^{-1}+\epsilon\right)+p_L\left(-x_k^{-1}-\epsilon\right)\right]=-\pi\period
\eeq
owing to \eqref{eq:sympx}.

It turns out that the function $p_L(x)$ coincides with the quasi-momentum computed in \cite{SV} (upon setting $\theta =a/2$). To see this, let us rewrite \eqref{eq:defofquasimomentum} using the identity,
\beq
\frac{1+x^2}{1-x^2}\frac{1}{(x-y)\left(1+\frac{1}{xy}\right)}=\frac{x^2}{1-x^2}\left(\frac{1}{x-y}+\frac{1}{x+\frac{1}{y}}-\frac{1}{x}\right)\period
\eeq
We then get
\beq\label{eq:prewriting}
p_L(x)=-\frac{a}{2}\frac{x^2+1}{x^2-1} +\frac{x}{x^2-1}\frac{L}{g}-\frac{x^2}{x^2-1}\sum_{k=1}^{2L}\frac{1}{x-x_k}\comma
\eeq
where we defined $x_k $ with $k>L$ as
\beq\label{eq:deflargek}
x_k \equiv -\frac{1}{x_{k-L}} \qquad (k>L)\period
\eeq 
The expression \eqref{eq:prewriting} coincides\fn{Precisely speaking, there is a small difference from (3.13) in \cite{SV}: In their case, the number of roots is $2L+1$ whereas it is $2L$ in our case, even after we doubled the number of roots by \eqref{eq:deflargek}. However, this difference does not affect the leading semiclassical answer.} with the definition of the quasi-momentum (3.13) in \cite{SV} if we take into account the fact that the distribution of $x_k$'s in \cite{SV} are symmetric under the transformation $x\to -1/x$. Furthermore, using \eqref{eq:deflargek}, the saddle-point equation can be re-expressed as
\beq
\frac{1}{2}\left[p_L(x_k+\epsilon)+p_L(x_k-\epsilon)\right]=\begin{cases}\pi \qquad &1\leq k\leq L\\-\pi\qquad &L+1\leq k \leq 2L\end{cases}\comma
\eeq
and it agrees with (3.15) in \cite{SV} (after appropriate reordering of $x_k$'s). These two agreements guarantee that our $p(x)$ has the same analytic properties as the quasi-momentum in \cite{SV}, which uniquely specify the function. We thus conclude that the two functions must be the same. 

Using the quasi-momentum $p_L(x)$, we can also express the semi-classical limit of $Q_L(x)$. By taking the saddle-point of the integral expression \eqref{eq:qlintegralexpression}, we obtain
\beq\label{eq:QLxsemiclassical}
Q_L(x) \sim g^{L}\prod_{k=1}^{L}(x-x_k)\left(1+\frac{1}{x x_k}\right)\comma
\eeq
where $x_k$'s are the saddle-point values of the integration variables, which satisfy \eqref{eq:saddlepoint}. Using the definition of $p_L(x)$, we can also rewrite \eqref{eq:QLxsemiclassical} as
\beq\label{eq:QLxsemiclassical2}
Q_L(x)\sim g^{L}\exp \left[-g\int^{u(x)} du^{\prime} \left(p_L(u^{\prime})+\frac{a}{2g}\frac{u^{\prime}}{\sqrt{(u^{\prime})^2-4g^2}}\right) \right]\comma
\eeq
where we introduced the rapidity variable $u$ defined by $u\equiv g (x+1/x)$.

Given the match of the quasi-momentum, we can follow the argument of \cite{SV} and show that the semi-classical limit of our matrix model correctly reproduces the Bremsstrahlung function computed from classical string. More interesting and challenging would be to compute the semi-classical limit of the structure constants using the integral representation \eqref{eq:integraltopologicalcor} and the asymptotic formula for $Q_L$ \eqref{eq:QLxsemiclassical2}. We leave this for future investigation.


\section{Conclusion\label{sec:conclusion}}
In this paper, we computed a class of correlation functions on the $1/8$ BPS Wilson loop by relating them to the area derivatives of the expectation value of the Wilson loop. When restricted to the $1/2$ BPS loop, the results provide infinitely many defect-CFT data. As a byproduct, we also obtained some of finite-$N$ generalizations of the generalized Bremsstrahlung function.

Let us end this paper by mentioning several future directions worth exploring: Firstly, it would be interesting to generalize our analysis to include operators outside the Wilson loop. In the absence of insertions on the loop, such correlators were computed in \cite{GP1,GP2} using the relation to 2d YM. Combining their results with our method, it should be possible to compute the correlators involving both types of operators. Work in that direction is in progress \cite{InProgress}.
Once such correlators are obtained, one can try to numerically solve the defect CFT bootstrap equation \cite{LM} using these topological correlators as inputs.

Another interesting direction is to apply our method to other theories, in particular to $\mathcal{N}=2$ superconformal theories in four dimensions, for which the Bremsstrahlung function was recently studied in \cite{FGK}. Having exact correlators for these theories would help us understand their holographic duals, including the dual of the Veneziano limit of $\mathcal{N}=2$ superconformal QCD \cite{GPR}.

At large $N$, we have shown that the correlators are expressed in terms of simple integrals. A challenge for the integrability community is to reproduce them from integrability. In the hexagon approach to the structure constants \cite{BKV,KK2}, the results are given by a sum over the number of particles. At first few orders at weak coupling where the sum truncates, it is not so hard to reproduce our results \cite{KK,Komatsu}. A question is whether one can resum the series and get the full results. In many respects, the topological correlators on the Wilson loop would provide an ideal playground for the hexagon approach; one can try to develop resummation techniques, fix potential subtleties (if any), and compute nonplanar corrections \cite{Handle,EJPS}.

Lastly, the appearance of the $Q$-functions in our large-$N$ results suggests deep relation between localization and the Quantum Spectral Curve. It is particularly intriguing that there is a one-to-one correspondence between the multiplication of the $Q$-functions and the operator product expansion of the topological correlators. A similar observation was recently made in \cite{CGL} in a slightly different context: They found that the correlators on the Wilson loop in the so-called ladders limit \cite{ladders}, which can be computed by resumming the ladder diagrams \cite{KKKN}, simplify greatly when expressed in terms of the $Q$-functions of the quantum spectral curve. Exploring such a  connection might give us insights into the gauge-theory origin of the Quantum Spectral Curve.
 
\section*{Acknowledgement}
We thank N.~Gromov, P.~Liendo, C.~Meneghelli and J.H.H.~Perk for useful discussions and comments. SK would like to thank N.~Kiryu for discussions on related topics. The work of SG is supported in part by the US NSF under Grant No.~PHY-1620542. The work of SK is supported by DOE grant number DE-SC0009988.
\appendix
\section{Explicit results for operators with $L\leq 3$\label{ap:explicit}}
In this Appendix we collect some explicit results for 2-point and 3-point functions of operators with $L\leq 3$. We restrict for simplicity to 
the case of the 1/2-BPS loop. In terms of the area-derivatives of the Wilson loop expectation value, one gets for the 2-point functions

{\footnotesize
\begin{eqnarray}
&&\langle \!\langle \NO{\tilde \Phi} \NO{\tilde \Phi} \rangle\! \rangle = 
\frac{\mathcal{W}^{(2)}}{\mathcal{W}}\\
&&\langle\! \langle \NO{\tilde \Phi^2} \NO{\tilde \Phi^2} \rangle \!\rangle =
\frac{\mathcal{W}\,\mathcal{W}^{(4)}-(\mathcal{W}^{(2)})^2}{(\mathcal{W})^2}\\
&&\langle\! \langle \NO{\tilde \Phi^3} \NO{\tilde \Phi^3} \rangle \!\rangle =
\frac{\mathcal{W}^{(2)} \,\mathcal{W}^{(6)}-(\mathcal{W}^{(4)})^2}{\mathcal{W}\, \mathcal{W}^{(2)}}
\end{eqnarray}
}
and for the 3-point functions
{\footnotesize
\begin{eqnarray}
&&\!\!\!\!\!\!\!\!\!\!\!\!\!\!\!\langle\! \langle \NO{\tilde \Phi^2} \NO{\tilde \Phi} \NO{\tilde \Phi}\rangle \rangle =\langle \langle \NO{\tilde \Phi^2} \NO{\tilde \Phi^2} \rangle\! \rangle \\
&&\!\!\!\!\!\!\!\!\!\!\!\!\!\!\!\langle\! \langle \NO{\tilde \Phi^2} \NO{\tilde \Phi^2} \NO{\tilde \Phi^2}\rangle\! \rangle =  
\frac{2 (\mathcal{W}^{(2)})^3-3 \mathcal{W}^{(2)} \mathcal{W}^{(4)} \mathcal{W}+\mathcal{W}^{(6)} (\mathcal{W})^2}{(\mathcal{W})^3}
 \\ 
&&\!\!\!\!\!\!\!\!\!\!\!\!\!\!\!\langle\! \langle \NO{\tilde \Phi^3} \NO{\tilde \Phi^2} \NO{\tilde \Phi^1}\rangle\! \rangle = 
\langle\! \langle \NO{\tilde \Phi^3} \NO{\tilde \Phi^3} \rangle\! \rangle \\
&&\!\!\!\!\!\!\!\!\!\!\!\!\!\!\! \langle\! \langle \NO{\tilde \Phi^3} \NO{\tilde \Phi^3} \NO{\tilde \Phi^2}\rangle\! \rangle = 
\frac{-(\mathcal{W}^{(2)})^3 \mathcal{W}^{(6)}+(\mathcal{W}^{(2)})^2 (\left(\mathcal{W}^{(4)})^2+\mathcal{W}^{(8)} \mathcal{W}\right)
-2 \mathcal{W}^{(2)} \mathcal{W}^{(4)} \mathcal{W}^{(6)} \mathcal{W}+(\mathcal{W}^{(4)})^3 \mathcal{W}}{(\mathcal{W}^{(2)})^2 (\mathcal{W})^2}
\,.
\end{eqnarray}
}

\noindent Here $\mathcal{W}\equiv <\mathcal{W}>|_{A=2\pi}$ and $ \mathcal{W}^{(k)}\equiv \frac{\partial^k}{\partial A^k}<\mathcal{W}>|_{A=2\pi}$ 
(similar expressions hold for the general 1/8-BPS loop, but they also involve derivatives of odd order). 
Using the Wilson loop expectation value (\ref{W-finite-N}), one can obtain in a straightforward 
way the explicit finite $N$ results in terms of Laguerre polynomials, but the expressions are rather lengthy and we do not report them here. 
In the planar large $N$ limit, the above correlators can be expressed in terms of Bessel functions as

{\footnotesize
\begin{eqnarray}
&&\langle \!\langle \NO{\tilde \Phi} \NO{\tilde \Phi} \rangle\! \rangle = 
-\frac{\sqrt{\lambda } I_2\left(\sqrt{\lambda }\right)}{4 \pi ^2 I_1\left(\sqrt{\lambda }\right)}\\
&&\langle\! \langle \NO{\tilde \Phi^2} \NO{\tilde \Phi^2} \rangle \!\rangle = \frac{3 \lambda }{16 \pi ^4}-\frac{\lambda  I_0\left(\sqrt{\lambda }\right){}^2}{16 \pi ^4 I_1\left(\sqrt{\lambda }\right){}^2}-\frac{\sqrt{\lambda } I_0\left(\sqrt{\lambda }\right)}{2 \pi ^4 I_1\left(\sqrt{\lambda }\right)}+\frac{5}{4 \pi ^4} \\
&& \langle \!\langle \NO{\tilde \Phi^3} \NO{\tilde \Phi^3} \rangle \!\rangle = \cr 
&& = -\frac{3 \sqrt{\lambda } (5 \lambda +72) I_0\left(\sqrt{\lambda }\right){}^2}{64 \pi ^6 I_1\left(\sqrt{\lambda }\right) I_2\left(\sqrt{\lambda }\right)}+\frac{3 (13 \lambda +144) I_0\left(\sqrt{\lambda }\right)}{32 \pi ^6 \left(I_0\left(\sqrt{\lambda }\right)-\frac{2 I_1\left(\sqrt{\lambda }\right)}{\sqrt{\lambda }}\right)}-\frac{3 (\lambda  (32-3 \lambda )+288) I_1\left(\sqrt{\lambda }\right)}{64 \pi ^6 \sqrt{\lambda } \left(I_0\left(\sqrt{\lambda }\right)-\frac{2 I_1\left(\sqrt{\lambda }\right)}{\sqrt{\lambda }}\right)}
\end{eqnarray}
}
and for the 3-point functions:
{\footnotesize
\begin{eqnarray}
&&\langle\! \langle \NO{\tilde \Phi^2} \NO{\tilde \Phi} \NO{\tilde \Phi}\rangle \rangle =\langle \langle \NO{\tilde \Phi^2} \NO{\tilde \Phi^2} \rangle\! \rangle \\
&& \langle\! \langle \NO{\tilde \Phi^2} \NO{\tilde \Phi^2} \NO{\tilde \Phi^2}\rangle\! \rangle =  -\frac{\lambda ^{3/2} I_0\left(\sqrt{\lambda }\right){}^3}{32 \pi ^6 I_1\left(\sqrt{\lambda }\right){}^3}+\frac{51 \lambda }{32 \pi ^6}-\frac{3 \lambda  I_0\left(\sqrt{\lambda }\right){}^2}{8 \pi ^6 I_1\left(\sqrt{\lambda }\right){}^2}-\frac{3 \sqrt{\lambda } (\lambda +40) I_0\left(\sqrt{\lambda }\right)}{32 \pi ^6 I_1\left(\sqrt{\lambda }\right)}+\frac{37}{4 \pi ^6} \\ 
&& \langle\! \langle \NO{\tilde \Phi^3} \NO{\tilde \Phi^2} \NO{\tilde \Phi^1}\rangle\! \rangle = 
\langle\! \langle \NO{\tilde \Phi^3} \NO{\tilde \Phi^3} \rangle\! \rangle \\
&& \langle\! \langle \NO{\tilde \Phi^3} \NO{\tilde \Phi^3} \NO{\tilde \Phi^2}\rangle\! \rangle = 
 -\frac{3 \lambda  (5 \lambda +72) I_0\left(\sqrt{\lambda }\right){}^4}{256 \pi ^8 I_1\left(\sqrt{\lambda }\right){}^2 I_2\left(\sqrt{\lambda }\right){}^2}
-\frac{3 \sqrt{\lambda } (127 \lambda +1920) I_0\left(\sqrt{\lambda }\right){}^3}{128 \pi ^8 I_1\left(\sqrt{\lambda }\right) I_2\left(\sqrt{\lambda }\right){}^2}
+\frac{3 (\lambda  (2 \lambda +579)+6192) I_0\left(\sqrt{\lambda }\right){}^2}{64 \pi ^8 I_2\left(\sqrt{\lambda }\right){}^2}\cr
&&+\frac{3 (\lambda  (5 \lambda -757)-6336) I_1\left(\sqrt{\lambda }\right) I_0\left(\sqrt{\lambda }\right)}{32 \pi ^8 \sqrt{\lambda } \left(I_0\left(\sqrt{\lambda }\right)-\frac{2 I_1\left(\sqrt{\lambda }\right)}{\sqrt{\lambda }}\right){}^2}
+\frac{3 (\lambda  (\lambda  (9 \lambda -112)+4960)+34176) I_1\left(\sqrt{\lambda }\right){}^2}{256 \pi ^8 \lambda  I_2\left(\sqrt{\lambda }\right){}^2}\,.
\end{eqnarray}
}
\subsection{Generalized Bremsstrahlung}
Let us also list the first few results for the generalized Bremsstrahlung function, focusing on the case $\theta=0$ given by eq.~(\ref{BLzero}). Using the 
same notation as above, the $L\le 2$ results in terms of area-derivatives of the Wilson loop expectation value read

{\footnotesize
\begin{eqnarray}
&&B_{L=0}(0) =-\frac{\mathcal{W}^{(2)}}{\mathcal{W}}  \\
&&B_{L=1}(0) = \frac{2 \mathcal{W}^{(2)}}{\mathcal{W}}-\frac{\mathcal{W}^{(4)}}{\mathcal{W}^{(2)}}\end{eqnarray}
}
Plugging in (\ref{W-finite-N}), one can find the explicit finite $N$ results. For instance, we obtain
{\footnotesize
\begin{eqnarray}
&&B_{L=0}(0) =\frac{\lambda}{16\pi^2N} \left(1+\frac{2 L_{N-2}^2\left(-\frac{\lambda }{4 N}\right)}{L_{N-1}^1\left(-\frac{\lambda }{4 N}\right)}\right)\\
&&B_{L=1}(0) =\frac{\lambda}{16\pi^2N} 
\left(1-\frac{4 L_{N-2}^2\left(-\frac{\lambda }{4 N}\right)}{L_{N-1}^1\left(-\frac{\lambda }{4 N}\right)}+\frac{6 \left(2 L_{N-3}^3\left(-\frac{\lambda }{4 N}\right)+L_{N-2}^2\left(-\frac{\lambda }{4 N}\right)\right)}{2 L_{N-2}^2\left(-\frac{\lambda }{4 N}\right)+L_{N-1}^1\left(-\frac{\lambda }{4 N}\right)}\right)\,.
\end{eqnarray}
}

\end{document}